\documentclass[acmsmall]{acmart}

\usepackage{booktabs}
\usepackage{epigraph}
\usepackage{multicol}
\usepackage{subcaption}
\usepackage{hyphenat}
\usepackage{xspace}
\usepackage{xcolor}
\usepackage{url}

\usepackage[show]{chato-notes} 

\newcommand{\hide}[1]{}

\newcommand{\xhdr}[1]{\vspace{1.7mm}\noindent{{\bf #1.}}}

\newcommand{\Secref}[1]{Sec.~\ref{#1}}

\newcommand{\Tabref}[1]{Table~\ref{#1}}
\newcommand{\Figref}[1]{Fig.~\ref{#1}}

\newcommand{\ie}{{i.e.}\xspace}
\newcommand{\eg}{{e.g.}\xspace}
\newcommand{\cf}{{cf.}\xspace}
\newcommand{\etal}{{et al.}\xspace}
\newcommand{\vs}{{vs.}\xspace}

\def\BibTeX{{\rm B\kern-.05em{\sc i\kern-.025em b}\kern-.08emT\kern-.1667em\lower.7ex\hbox{E}\kern-.125emX}}

\begin{document}

\setcopyright{acmlicensed}
\acmJournal{PACMHCI}
\acmYear{2019} \acmVolume{3} \acmNumber{CSCW} \acmArticle{45} \acmMonth{11} \acmPrice{15.00}\acmDOI{10.1145/3359147}

\copyrightyear{2019}

\received{April 2019} 
\received[revised]{June 2019}
\received[accepted]{August 2019}

\title[Causal Effects of Brevity on Style and Success in Social Media]{Causal Effects of Brevity on Style and Success\\in Social Media}

\author{Kristina Gligori\'c}
\affiliation{%
  \institution{EPFL}
  \city{Lausanne, Switzerland}
}
\email{kristina.gligoric@epfl.ch}

\author{Ashton Anderson}
\affiliation{%
  \institution{University of Toronto}
    \city{Toronto, ON, Canada}
}
\email{ashton@cs.toronto.edu}

\author{Robert West}
\affiliation{%
  \institution{EPFL}
  \city{Lausanne, Switzerland}
}
\email{robert.west@epfl.ch}

\renewcommand{\shortauthors}{Gligori\'c, Anderson, and West}

\begin{abstract}
In online communities, where billions of people strive to propagate their messages, understanding how wording affects success is of primary importance. In this work, we are interested in one particularly salient aspect of wording: brevity. What is the causal effect of brevity on message success? What are the linguistic traits of brevity? When is brevity beneficial, and when is it not?

Whereas most prior work has studied the effect of wording on style and success in observational setups, we conduct a controlled experiment, in which crowd workers shorten social media posts to prescribed target lengths and other crowd workers subsequently rate the original and shortened versions. This allows us to isolate the causal effect of brevity on the success of a message. We find that concise messages are on average more successful than the original messages up to a length reduction of 30--40\%. The optimal reduction is on average between 10\% and 20\%. The observed effect is robust across different subpopulations of raters and is the strongest for raters who visit social media on a daily basis. Finally, we discover unique linguistic and content traits of brevity and correlate them with the measured probability of success in order to distinguish effective from ineffective shortening strategies. Overall, our findings are important for developing a better understanding of the effect of brevity on the success of messages in online social media.
\end{abstract}

\keywords{brevity; conciseness; linguistic style;  Twitter; social media; microblogging; causal effects; experimental methods; crowdsourcing}

\maketitle

\section{Introduction}
\label{sec:intro}

\begin{flushright}
{\small
\textit{I didn't have time to write a short letter, so I wrote a long one instead.}\,---Blaise Pascal
}
\end{flushright}


\noindent
Being concise%
\footnote{We use the two words ``brief'' and ``concise'' interchangeably in this paper.}
when communicating has been encouraged throughout history. 
In the 1st century BC, Cicero said that ``Brevity is a great charm of eloquence,'' and in the 17th century, Shakespeare wrote in \emph{Hamlet} that ``Brevity is the soul of wit.''
German teachers commonly advise their students to respect the proverb \textit{``In der K\"urze liegt die W\"urze''} (``The spice is in the concise'') when writing essays, and researchers submitting papers to this conference have been instructed that ``Papers should report research thoroughly but succinctly: brevity is a virtue'' \cite{CSCW_cfp}.

The online world has followed this long tradition of promoting brevity. Short messages are well-suited to small screens, and images with few words in large text are often shared widely. Many social media platforms even enforce hard length constraints, epitomized by Twitter's signature character limit. Online messages are thus shaped by length constraints, with far\hyp reaching consequences: as McLuhan~\cite{mcluhan1994understanding} famously pointed out, ``The medium is the message,'' and the constraints that a medium imposes affect the audience not only through the content delivered over the medium, but also through the characteristics of the medium itself. Therefore, as social media continue to play an increasingly critical role in modern societies---influencing what we read, what we buy, and whom we elect into office---, understanding how constraints, especially brevity constraints, affect the effectiveness of social media messages has major societal as well as financial ramifications.

In November 2017, Twitter, one of the most prominent social media platforms, changed course by relaxing its hallmark brevity constraint, doubling the maximum tweet length from 140 characters to 280 characters.
Brevity's long and storied reputation, alongside this recent change in policy, raises questions:
How do brevity constraints affect messages? 
What precisely, if any, are the benefits of brevity?
Expressing one's thoughts in few words may be more difficult than doing so in many (as Blaise Pascal quips in the opening quote)---is brevity worth the effort?

\begin{figure*}
    \includegraphics[width = \textwidth]{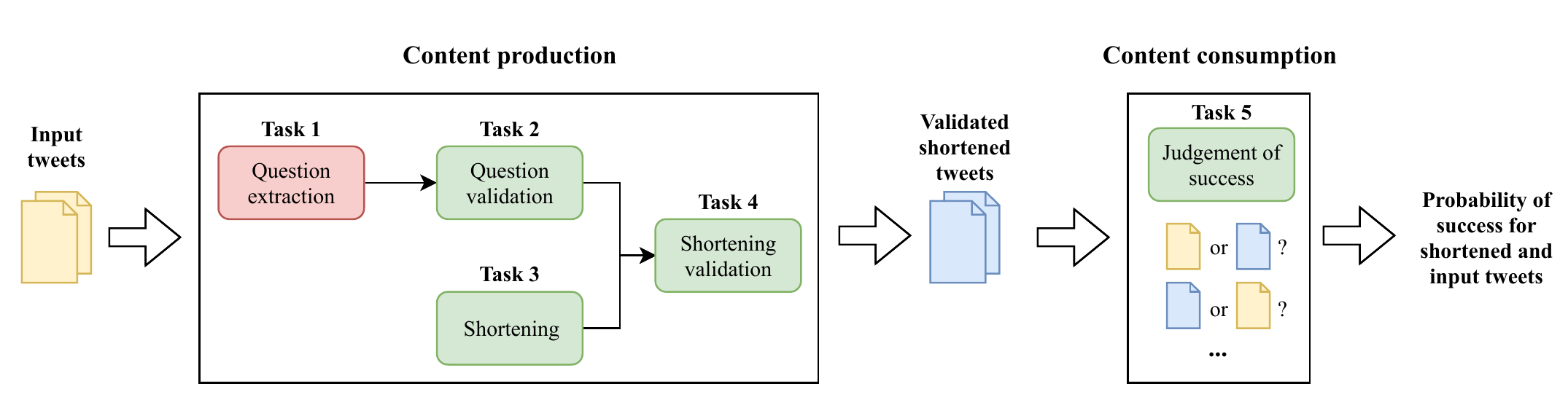}
    \caption{Schematic diagram of the experimental design. The experiment consists of two parts, designed to replicate the  \textit{production} and \textit{consumption} of textual content in online social media. The goal of the content production part (Tasks 1--4) is, for a given input tweet, to output shortened versions while preserving the meaning of the input tweet. In the content consumption part of the experiment (Task 5), we show participants pairs of tweets (a short tweet [treatment] and the corresponding long original tweet [control]) and ask which one is more likely to see more retweets. The output of this setup is binary votes (several per pair), based on which we compute the probability of success for each tweet version.}
    \label{fig:0}
\end{figure*}

\xhdr{Summary of contributions}
To the best of our knowledge, this is the first experimental study of the effect of brevity in online social media.

Starting from an original, long tweet of exactly 250 characters, we ask crowd workers to shorten it into 9 different lengths, while preserving all the essential information contained in the original tweet.
We then ask a separate group of crowd workers to compare each shortened version to the original tweet and indicate which one of the two---long or short---is of higher quality.%
\footnote{The input tweets together with the outputs from the experiment are available at \url{https://github.com/epfl-dlab/brevity}.}

If crowd workers made no mistakes, neither intentional nor unintentional, during shortening and rating, this setup would indeed meet the requirement of comparing the exact same information manifested at various levels of brevity.
Unfortunately, however, crowd workers are not perfect: they may make unintentional mistakes and may also be intentionally sloppy in order to minimize the effort spent per dollar earned, \eg, by simply cutting trailing characters in the shortening task, or voting randomly in the rating task.
Our full study design, depicted schematically in \Figref{fig:0}, is therefore more sophisticated.

Applying our experimental framework (\Secref{sec:experiment}), we collect short versions at 9 brevity levels for 60 original tweets, judged against the original tweets in a total of 27,000 binary votes.
Based on this dataset, using mixed quantitative and qualitative methods, we address three core \textbf{research questions:}

\begin{enumerate}
    \item[\textbf{RQ1:}] What are the causal effects of brevity on the success of social media content? (\Secref{sec:results})
    \item[\textbf{RQ2:}] What are the linguistic traits of brevity? (\Secref{sec:languagea})
    \item[\textbf{RQ3:}] When is brevity beneficial, and when is it not? (\Secref{sec:languageb})
\end{enumerate}

Regarding RQ1, we find that concise versions are on average more successful than the original messages up to a length reduction of 30--40\%, while the optimal reduction is on average between 10\% and 20\% (resulting length 211--215 characters, down from the original 250 characters). This effect is robust across different subpopulations of raters and is strongest for raters who visit social media on a daily basis.

With respect to RQ2, we find that the shortening process disproportionally preserves verbs and negations---parts of speech that carry essential information---in contrast to, \eg, articles and adverbs. The shortening also preserves affect and subjective perceptions, and the effect is strongest for negative emotions.

Addressing RQ3, we find initial evidence that it is effective to omit certain function words and to insert commas and full stops, presumably as this increases readability by structuring or splitting long sentences. Ineffective editing strategies include deleting hashtags as well as question and exclamation marks.

We begin by positioning our contributions in the context of related work (\Secref{sec:rel}), and conclude by discussing the importance and implications of our findings (\Secref{sec:discussion}).

\section{Related work}
\label{sec:rel}
\subsection{Observational studies}

We start our review of related work by discussing previous attempts to address our main research question, namely how brevity affects the success of social media content.

Online settings present many opportunities for measuring the benefits of brevity. For example, researchers can link the success of posts (\eg, the number of shares or likes) to linguistic features including length and wording.
It is tempting to study the benefits of brevity online in a straightforward fashion and correlate the length of social media posts with their success, as done by countless social media ``pundits'', who observe, \eg, that ``tweets containing less than 100 characters receive, on average, 17 percent higher engagement than longer tweets'' \cite{hootsuite} and conclude that users, therefore, should use fewer than 100 characters in their tweets.
Unfortunately, this na\"ive approach is fraught with confounding factors:
The length of a tweet may be correlated with its topic, and some topics may be inherently more shared, but due to their inherent attractiveness, not due to the length or brevity of the tweets users write about them.
Also, users who write better content in general might also write more concisely; so what causes more engagement in these users' tweets might be the overall quality of content, rather than the fact that the content is also concise.
Finally, posts written at certain times (\eg, during hectic lunch hours) might be both more concise as well as more interesting than posts written at other times (\eg, lengthy posts written out of boredom during a sleepless night).

To truly establish a causal link between brevity and success, one would need to compare posts that convey the exact same semantic information and differ only in the number of characters used to express the fixed semantic content in a specific lexical and syntactic surface form.
Some observational studies have striven to approximate this ideal goal by carefully controlling for confounding factors.
Tan \etal\ \cite{r9}
consider pairs of tweets that were posted by the same user and contain the same URL, but are differently worded otherwise, thus avoiding user- and certain topic\hyp related confounds.
Analyzing such pairs, they establish correlations between the wording and success (in terms of retweets) of messages, finding that the longer tweet within a pair is on average more successful, and concluding that being long and informative beats being concise.

More recently,
we
exploited the aforementioned relaxation of Twitter's length constraint from 140 to 280 characters as a natural experiment \cite{w1}:
before the switch, tweets of exactly or just under 140 characters were likely to have been squeezed to satisfy the length constraint, whereas this was no more the case after the switch, for tweets of the same length.
Hence, comparing tweets of exactly or just under 140 characters before \vs\ after the switch (while avoiding user- and certain topic\hyp related confounds by comparing only tweets written by the same user or containing the same hashtags), allowed
us
to approximate the effects of length constraints on the style and success of content.
As opposed to Tan \etal\ \cite{r9},
we found
that length constraints have a mild positive effect on message success (in terms of retweets).

In addition to these contradictory conclusions, and although the two respective methodologies constitute a big improvement over the above\hyp sketched na\"ive approach, another major problem remains:
fixing users, hashtags, or URLs cannot ultimately guarantee that the semantic content of two compared posts is exactly identical, and length may still be confounded with other factors such as the inherent attractiveness of a message.
In summary, prior research has not been able to offer a convincing answer to the question of the benefits of brevity on social media.

In order to overcome the methodological hurdle inherent in observational designs and to more closely approximate the ideal of comparing two messages---one long, one short---expressing the exact same semantic content, we adopt an experimental approach instead.

\subsection{Additional related work}

In addition to the prior work discussed above, our work draws from, and has implications for, many other research threads. The purpose of the present section is to position our work more broadly within prior research on the effect of wording and length on success. 

\xhdr{Effects of wording}
The question how to word a message in order to convey it most successfully is important across a number of contexts, \eg, in political campaigns, marketing slogans, or writing convincing manuscripts.
In the context of online communities, too, understanding what wording makes messages successful is of growing importance.
The question is often formulated as the task of predicting what makes textual content become popular \cite{berger2012makes,guerini2011exploring,lamprinidis2018predicting}.
In the specific case of Twitter, in addition to characterizing how language is used on the platform in general \cite{murthy2012towards,levinson2011long,hu2013dude,eisenstein2013bad}, researchers have investigated the correlation of linguistic signals with the propagation of tweets \cite{artzi2012predicting,bakshy2011everyone,r9,doi:10.1080/15534510.2016.1265582}.
Going beyond correlations, we add to the literature by designing and deploying an experimental, crowdsourcing\hyp based framework to establish a causal link between a particularly salient linguistic signal---brevity---and message success. 

\xhdr{Benefits of communicating concisely}
Philology, communication, education, and psychology scholars have investigated conciseness and its benefits in many different contexts  \cite{laib1990conciseness,vardi2000brevity,sloane2003say}. A company's weak financial performance correlates with less readable and less concise accounting reports \cite{melloni2017saying}. Successful marketing slogans are short. None of the other design elements (such as complexity of slogans, use of jingles, and use of rhymes) have an impact on slogan recall \cite{kohli2013you}. Conciseness is also preferred in spatial discourse \cite{daniel2004production}.

\xhdr{Creativity and constraints}
Perhaps counterintuitively, research across several fields, such as product design, process management, and education, suggests that the presence of a constraint (limited time, money, or some other resource) is beneficial \cite{i1,i2,i3,i4,i5}.
In particular, length constraints are often thought to have a positive impact on the quality of content. For instance, the conciseness of research paper titles correlates with how frequently a research paper is cited \cite{i8}. Research on the effect of constraints on creativity reveals the presence of two opposite effects---restricting freedom directs the search for solutions and degrades creativity, but simultaneously, having too much freedom can be paralyzing and detrimental. The trade-off with the best compromise between the two extremes occurs at a sweet spot with just the right amount of constraints \cite{i7}. (For a more complete account, we refer the reader to Onarheim and Biskjaer's survey on creativity constraints \cite{ii6}.) This interplay is a central open  question in creativity research, and it remains unclear how these complex phenomena reflect on the universe of online social media.

\xhdr{Crowdsourced experiments replicating online social media}
Past work that has also recruited crowd workers to produce content in simulations of online social media settings includes research on trolling \cite{Cheng:2017:ABT:2998181.2998213}.  
Experiments involving crowd workers have also been used to simulate the consumption of content, \eg, by Tan \etal \cite{r9}, who demonstrate that, when considering the majority vote among 39 workers, the accuracy in guessing which one of two messages had been shared more frequently was 73\%.

\xhdr{Tweet summarization}
Prior studies have defined the tweet generation problem within a summarization framework, mostly focusing on generating indicative summaries, e.g., generating tweets about news articles \cite{r7,r8,r10,r11}. For the largest part, these works have been limited to automated extractive summarization approaches and corresponding automated evaluation metrics. Finally, the idea of reducing the length of a tweet while maintaining its information content is closely related to rich work centered around the linguistic tasks of paraphrasing \cite{r5,prakash2016neural,yin2015convolutional}, reduction \cite{r1,r2}, and simplification \cite{r3,r4,r6}.

\xhdr{Information overload and attention budget}
Our findings also have implications for several phenomena related to information consumption (rather than production). Firstly, social media users suffer from information overload, being exposed to an endless flow of information online, far surpassing the rate at which they can process it \cite{o1gomez2014quantifying,o2bawden2009dark,o3jones2004information}. Secondly, social media users have a limited attention budget \cite{backstrom2011center,jiang2013optimally}, governing them in selecting a subset of users or topics to follow. In such scenarios, where users are exposed to amounts of information that severely surpass their capacity, brevity is particularly desirable. 

\xhdr{Perceptual fluency}
Finally, needless complexity negatively impacts raters' assessments of text, off\/line and online.
Wrapping familiar ideas in pretentious language has been shown to be interpreted as a sign of poor intelligence and low credibility \cite{oppenheimer2006consequences}.  Additional empirical confirmation that people prefer content that is easy to process comes from research on perceptual fluency \cite{dragojevic2016don}.

\section{Experimental design}
\label{sec:experiment}

To establish the causal effect of brevity on message success, we designed and conducted an online crowdsourced experiment. The experiment consists of two parts, designed to replicate the  \textit{production} and \textit{consumption} of textual content in online social media. The goal of the first part is, for a given input tweet, to output several shortened versions of the tweet, while preserving its meaning as much as possible. The goal of the second part is to estimate which version of a tweet is better, the original or a shorter one.

We start by describing the experimental design outlined in \Figref{fig:0}. We designed a series of crowdsourcing tasks with the purpose of obtaining pairs of tweets consisting of an original tweet and a concise version, and an estimate of which of these is perceived as more successful.
In this section, we first delineate our crowdsourcing framework (\Secref{sec:Crowdsourcing framework}) and then supply additional details on how we implemented this framework on Amazon Mechanical Turk (\Secref{sec:Implementation on Amazon Mechanical Turk}).

\subsection{Crowdsourcing framework}
\label{sec:Crowdsourcing framework}

\subsubsection{Content production}

In the first part of our experimental design, the goal is to extract accurate shortened versions of tweets. We accomplish this by
extracting comprehension questions for each tweet (Task~1),
validating the questions (Task~2),
asking crowd workers to shorten tweets without sacrificing essential information (Task~3),
and finally validating the shortened tweets with the comprehension questions (Task~4).
See \Tabref{tab:2} for an example of a tweet alongside shortened versions and comprehension questions.

\xhdr{Task 1}
In the question extraction task, our goal is to extract three comprehension questions that capture the essential message of the tweet (alongside two candidate answers per question), following these guidelines:
\begin{enumerate}
    \item The three questions should cover the most important information contained in the tweet.
    \item The answers to the questions must be contained in the tweet itself (but not necessarily in the exact same wording).  Also, answering the questions should require reading the tweet. (In other words, we do not formulate questions that can be answered using commonsense knowledge about the world alone.)
    \item Only one of the two candidate answers can be correct, according to the tweet.
\end{enumerate}

\xhdr{Task 2}
In the question validation task, we validate whether crowd workers can indeed answer the questions given the original tweet. 

\xhdr{Task 3}
In the shortening task, we ask each worker to shorten a tweet to a specific length. Workers are instructed that it is essential to not remove any important information in the process: ``The meaning of the original tweet must be maintained. It is important that you don't remove any important information by shortening the tweet!'' The workers are also provided with an example.

\xhdr{Task 4}
In the shortening validation task, we ensure that the meaning is preserved when shortening, by having workers answer the comprehension questions based solely on the short tweet and maintaining only those short tweets for which the answers are consistently correct. This task also lets us quantify to what extent it is even possible to reduce tweet length while maintaining meaning.

\subsubsection{Content consumption} The second part of the experimental pipeline is designed to replicate tweet consumption and determine for each short version whether it is better than the original tweet.

\xhdr{Task 5}
In the success judgment task, we show participants pairs of tweets, one treated (shortened) tweet and the control (original, long) tweet. We ask: ``Below, you are given two different tweets. Which one do you think will get more retweets?'' We deliberately do not ask which one they would rather retweet personally, as that question is more dependent on particular subjective opinions of individual participants.
Aggregating multiple votes for the same pair, we estimate the probability of success for each shortened version.
Additionally, as an attention check for this task, we repurpose the previously extracted questions and have participants answer them to exclude votes from inattentive participants when estimating probabilities of success.

\subsection{Implementation on Amazon Mechanical Turk}
\label{sec:Implementation on Amazon Mechanical Turk}

Building on the fact that regular crowd workers can reduce the length of text by up to 70\% without cutting any major content \cite{p1} and can accurately estimate which of two messages in a pair (for a fixed topic and user) will be shared more frequently \cite{r9}, we deployed our experiment on the Amazon Mechanical Turk crowdsourcing platform (AMT).

In the remainder of this section, we describe how we have implemented the above experimental framework (Tasks 2--5) on Amazon Mechanical Turk. The experiment was carried out over 16 days.

\xhdr{Participants}
Since the tasks require comprehension and production of text in English, participants were restricted to those residing in the United States, Canada, or the United Kingdom. To ensure high-quality answers, we admitted only workers with approval rates greater than 98\% and with more than 200 previously approved tasks. To avoid demand biases, the participants were not informed of the purpose of the experiment. We also gathered information about Turkers' demographics (gender, age, education), online presence (the frequency of visiting social media and whether they hold a Twitter account), and reading habits (the frequency of reading books). We targeted a pay rate of \$9 per hour. Summary statistics including the completion time and pay rate for each task are presented in \Tabref{tab:6}.

\begin{table}[b]
\footnotesize
  \caption{Summary statistics of crowdsourcing tasks.}
  \label{tab:6}
\begin{tabular}{lcc}
\toprule
 & \textbf{Price per task} & \textbf{Median completion time}\\
\midrule
\textbf{Task 1} & \multicolumn{2}{c}{\textit{Task performed by authors, not crowd workers}} \\
\textbf{Task 2} & \$0.10 & 36 seconds\\
\textbf{Task 3} & \$0.10 & 48 seconds \\
\textbf{Task 4} & \$0.10 & 43 seconds \\
\textbf{Task 5} & \$0.30 & 103 seconds \\
 \end{tabular}
\end{table}

\xhdr{Input tweets}
In order not to overfit to idiosyncrasies of particular authors, we sampled tweets from distinct users, based on a dataset containing full timelines of users present in the 1\% sample that Twitter supplies via its Spritzer API, for the posting time between April and June 2017. We took users with a medium number of followers (between the first and third quartiles, \ie, between 4 and 65 followers \cite{Myers:2014:INS:2567948.2576939}), which avoided tweets from highly influential users who might be producing content of atypically high quality.
To have tight control, each input tweet was exactly 250 characters long. Such tweets are unlikely to have been optimized to fit the 280\hyp character constraint imposed by Twitter. We chose only tweets written in English. Also, since we are interested in textual content independently produced by the user, we excluded tweets that contain URLs, @ mentions of other users or are replies to other tweets. All input tweets were posted between December 2017 and August 2018.
We randomly sampled 60 tweets adhering to these inclusion criteria.

We selected the number of input tweets and the number of raters per pair using power analysis, as follows. We generated votes as a function of reduced length using a logistic regression model, varying the model parameters, the number of input tweets, the number of participants per pair, and the level of additive noise. We then studied with what number of input tweets and number of participants we could recover the slope of the logistic regression model, i.e., reject the null hypothesis that the slope is zero. Given that already with 60 input tweets and 50 votes per tweet we could reject the null hypothesis when the value of the slope was as small as $0.05$, we decided to chose those parameters.

\xhdr{Task 1}
To be sure that the guidelines were followed, we manually extracted the comprehension questions ourselves (hence the different color of Task~1 in \Figref{fig:0}).

\xhdr{Task 2}
In this task, we assigned 60 workers to answer the three manually extracted comprehension questions, one worker for each original tweet. We require that the three questions are answered correctly by the worker they were assigned to. All manually extracted questions were correctly answered in this task and accepted as valid.

\xhdr{Task 3}
Our main research goal is to measure the causal effect of length constraints on message success. The treatment in our experimental design is the length constraint that we impose on tweets. For each worker assigned to an input tweet, we randomly assign a length constraint to that worker such that each input tweet is shortened multiple times, once for each target length.

Workers were asked to shorten each tweet to 8 non\hyp overlapping length buckets, each 5 characters wide. The buckets are ranging from 10\% to 90\% of the original length (250 characters), \ie, $[36,40], [61,65],$ $ \ldots,[211,215]$ characters. To ensure this is done properly, it was not possible to submit work unless the edited tweet was strictly inside the requested range. In order to study the effect on a fine-grained level, we use a full factorial design: all tweets are exposed to all treatments, that is 60 tweets shortened to 8 non\hyp overlapping lengths each are compared against the corresponding original tweets.
Task~3 was completed by 99 distinct workers.

To distinguish the benefits of brevity from the benefits of simple editing (such as correcting typos), we introduced a baseline length that lies at least one and at most five characters below the original length. Each original tweet was also compared to the baseline version. If there were no additional benefits of brevity beyond simple editing, we would expect length constraints to perform about equally with this baseline.

\begin{figure}
    \begin{minipage}{.3\columnwidth}
        \centering
        \includegraphics[width=\textwidth]{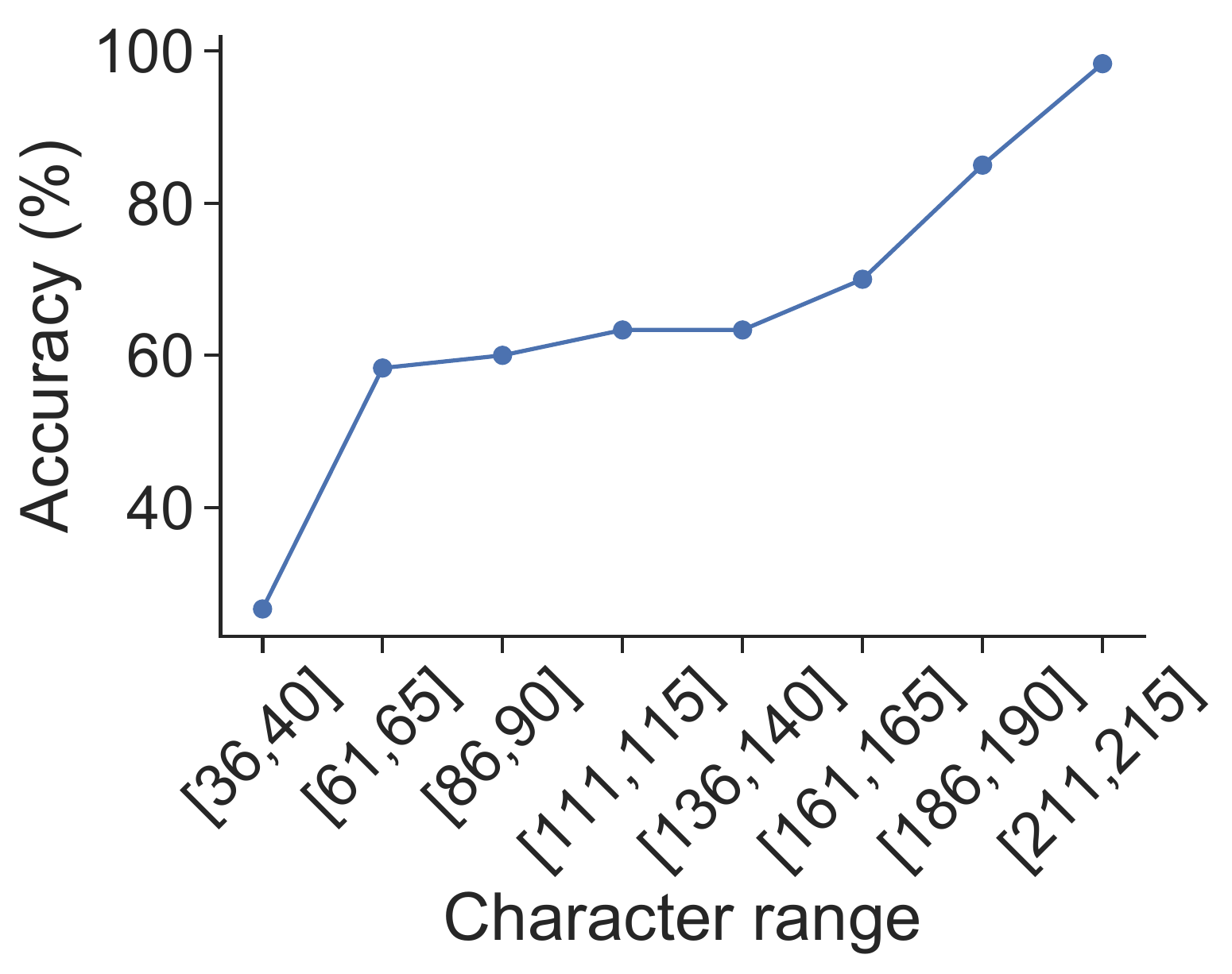}
        \subcaption{Accuracy on Task 4}
        \label{fig:9}
    \end{minipage}
    \hfill
    \begin{minipage}[]{.3\columnwidth}
        \centering
        \includegraphics[width=\textwidth]{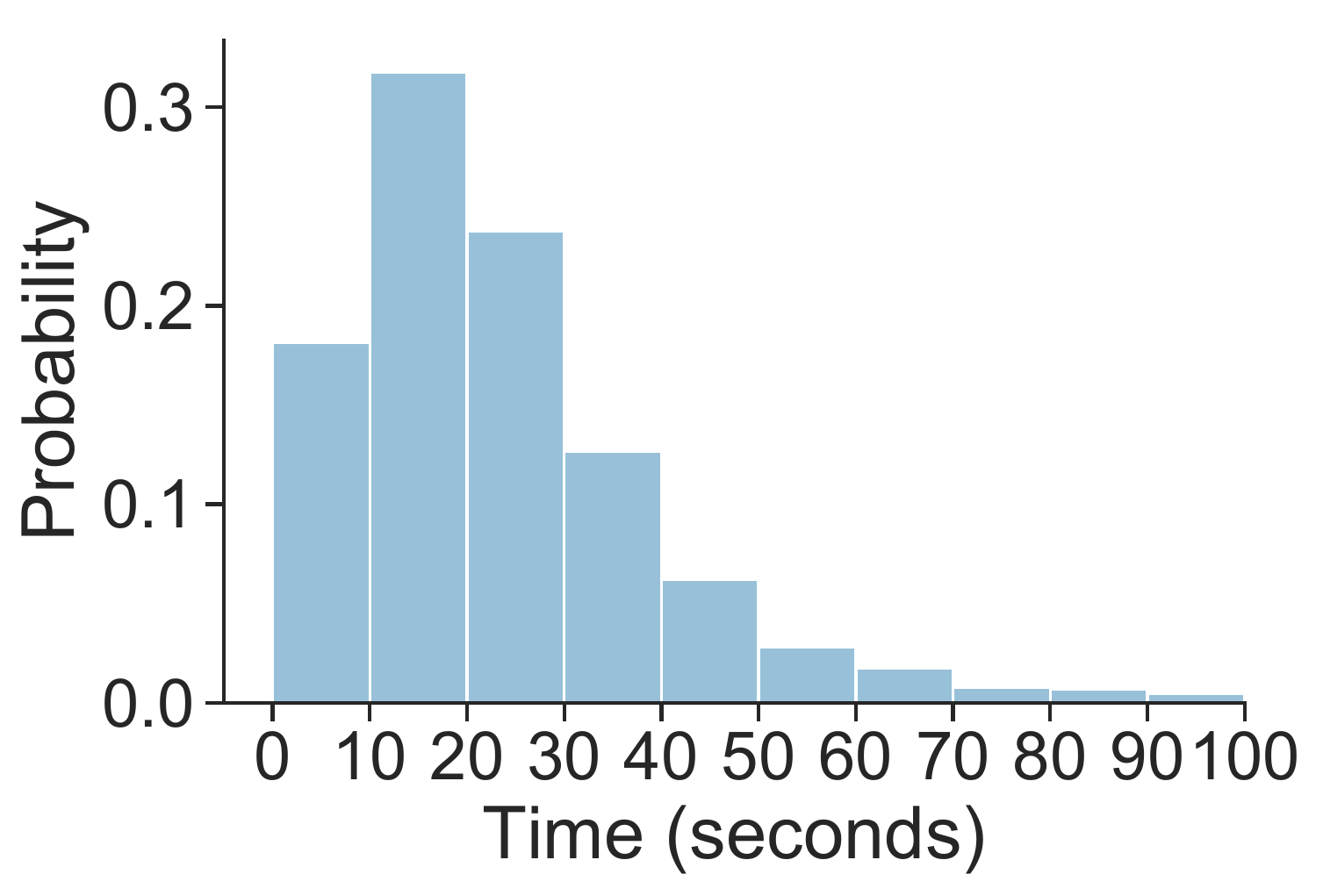}
        \subcaption{Work time}
        \label{fig:7}
    \end{minipage}
    \hfill
    \begin{minipage}{.3\columnwidth}
        \centering
        \includegraphics[width=\textwidth]{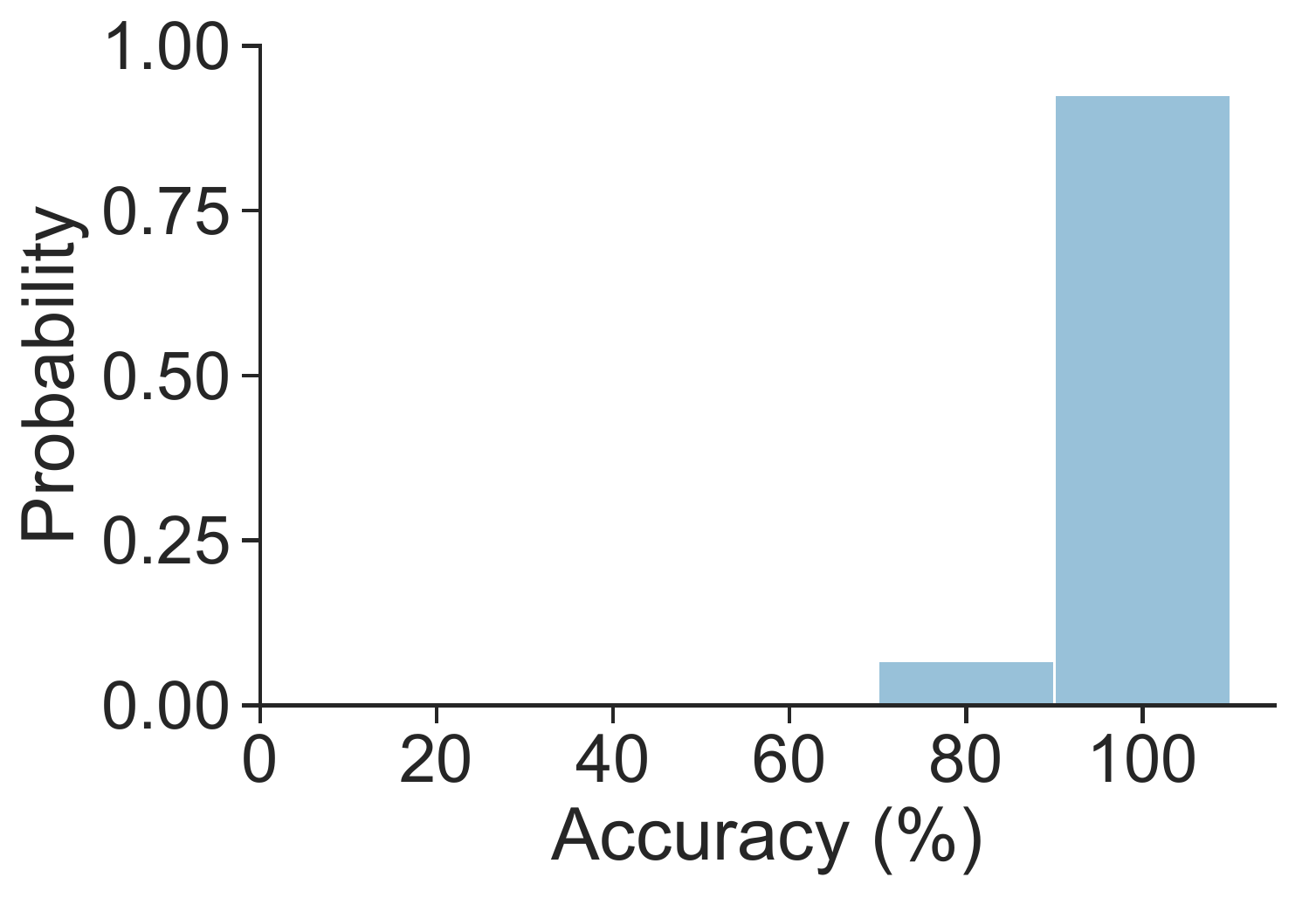}
        \subcaption{Attention check accuracy}
        \label{fig:8}
    \end{minipage}
    
    \caption{
    \textbf{(a)}~Worker accuracy when answering validation questions about original tweets based on short versions only (Task 4). The accuracy increases as the shortening becomes less drastic. When the message is reduced to only up to 20\% of the original length (36--40 characters), it becomes essentially impossible to maintain the meaning of the message.
    \textbf{(b)}~Histogram of work time in the success judgment task (Task 5), in seconds per pair (median: 20 seconds).
    \textbf{(c)}~Histogram of worker accuracy when answering comprehension questions (Task 5).
    }
\end{figure}

\xhdr{Task 4}
For each shortened version, we assigned a worker who answered the three comprehension questions based on the shortened tweet. Workers performing Tasks 2 and 3 could not participate, as they had already been exposed to the original tweets and had thus potentially more information than contained in the shortened version.
We evaluated the accuracy when answering questions, revealing that it is possible to reduce the length while maintaining the critical information down to 20\% of the original length (\Figref{fig:9}).
We require that all three comprehension questions were correctly answered (except for the largest level of shortening, \ie, 10--20\% of the original length). If they were not, the tweet was discarded and the shortening was performed again, by another worker, to ensure that we gather all validated shortened versions for all tweets. Note that, from here on, in the experiment and in the analysis we only consider validated shortened tweets where all questions were answered correctly, except for the largest level of shortening (10--20\% of the original length), where it becomes essentially impossible to maintain the meaning. Task~4 was performed by 109 distinct workers.

\xhdr{Task 5}
For each pairwise comparison in Task 5 (60 tweets in 9 conditions being compared to the original tweet), we obtained votes from 50 unique workers, resulting in a total number of 27,000 binary votes. Workers submitted five binary votes as part of one assignment, for randomly assigned pairwise comparisons. When presented to the participants, the order of the treated and untreated tweets within a pair was randomized. Workers also answered five comprehension questions, one for each tweet. Workers who participated in the first part of the experimental pipeline (content production) could not participate, as they might be biased in their preference, having shortened the tweets themselves.

To gauge the quality and reliability of the gathered data in Task 5, we examine how much time participants took to decide on a single pair of tweets (\Figref{fig:7}; at a median of 20 seconds, workers took significant time to decide which version of a tweet is better, a sign that they performed the task thoroughly) and the accuracy of answering comprehension questions (\Figref{fig:8}; as we can see, workers answered nearly perfectly). To exclude potentially malicious or inattentive workers in Task 5, we discard votes from workers with low accuracy on attention checks (less than 80\% on average), workers with an average work time of less than 10 seconds per pair, and workers who consistently chose the left or the right tweet in a pair. Task~5 was performed by 623 workers.

\section{RQ1: Benefits of brevity}
\label{sec:results}

\begin{table*}[b]
\scriptsize
  \caption{Example of baseline shortening, where the tweet is shortened by one to five characters. Edits correcting the spelling, grammar, and punctuation are emphasized in bold (78\% of workers voted in favor of the baseline version).}
  \label{tab:1}
\begin{tabular}{ll}
    \toprule
    \textbf{Original tweet} & When we as Black \textbf{American} STOP apologizing for being Black\textbf{. Stop} assimilating to White culture and start  \\
    & respecting that it is Black culture that White People \textbf{kling} to, imitate and pattern \textbf{,} then \textbf{an} only then will we  \\
    & as Black \textbf{American} \textbf{will} be alright \\
    \hline
    \textbf{Baseline tweet} & When we as Black  \textbf{Americans} STOP apologizing for being Black\textbf{, stop} assimilating to White culture\textbf{,} and   \\
    &start respecting that it is Black culture that White People \textbf{cling} to, imitate and pattern\textbf{,} then \textbf{and} only then will we \\
    &as Black \textbf{Americans} be alright\textbf{.} \\
  \end{tabular}
\end{table*}

\begin{table*}[b]
\scriptsize
  \caption{Example of an input tweet alongside extracted questions with possible answers (the correct answers are underlined), shortened versions, and the probability of success of the shortened version (``pr.\ succ.''), \ie, the fraction of workers who voted for the shortened version over the input tweet. \textcolor{teal}{Colored} tweets are versions where the short version was preferred over the original one on average.}
  \label{tab:2}
\begin{tabular}{clc}
    \toprule
    \textbf{Input tweet} &\textbf{And when I finally shared that secret, that was when I was able to start healing. Don't get me } & \\ 
    &\textbf{wrong, addiction is always that, it's not always easy. However, whenever I feel weak, I put on} & \\
    &\textbf{Buffy. There's so much that I can connect with in that show.} & \\
    \midrule
    \textbf{Extracted} & When was the person able to start healing? (\underline{Once they shared the secret} / Once they forgot) & \\
    \textbf{questions} & Is addiction easy according to the tweet? (Yes / \underline{No}) & \\
     & Why do they watch Buffy? (\underline{They can connect with that show} / It's about addiction) & \\
    \midrule
    \textbf{Condition} &\textbf{Shortened tweet}&\textbf{pr.\ succ.} \\
    \midrule
    
    \textbf{Baseline} & And when I finally shared that secret, that was when I was able to start healing. Don't get me wrong, & 0.46 \\
    & addiction is always that, it's not always easy. However, whenever I feel weak, I put on Buffy. There's & \\
    & so much I can connect with in that show. & \\

    \textbf{80-90\%}&\textcolor{teal}{Don't misunderstand me, addiction is never easy. However, when I finally shared that secret, I was able }& 0.75\\
    &\textcolor{teal}{to start healing. Whenever I feel weak, I put on Buffy. There's so much that I can connect with in that show.}&\\

    \textbf{70-80\%} &\textcolor{teal}{When I shared that secret, I was able to start to heal. addiction is not always easy. However, whenever I }& 0.69\\
    &\textcolor{teal}{feel weak, I put on Buffy. There's so much that I can connect with in that show.}& \\

    \textbf{60-70\%} &\textcolor{teal}{When I shared that secret, I started healing. Don't get me wrong, it's not always easy. When I feel weak, }& 0.72\\
    &\textcolor{teal}{I put on Buffy. So much that I connect with in that show}& \\

    \textbf{50-60\%} &\textcolor{teal}{When I shared that secret I could start healing. Addiction is rarely easy. When I feel weak I put on Buffy, } & 0.63\\
    &\textcolor{teal}{I really connect to that show.} & \\

    \textbf{40-50\%} &Addiction is not always easy. However, whenever I feel weak, I put on Buffy. Because I can connect with & 0.39\\
    & that show. & \\

    \textbf{30-40\%} &\textcolor{teal}{Admitting addiction started my healing. And when really bad, watching Buffy is a blessing!}& 0.54\\

    \textbf{20-30\%} &Sharing my secret healed me. I can connect with a lot on Buffy.& 0.32\\

    \textbf{10-20\%} &Addiction can heal with shared secrets.& 0.34\\
  \end{tabular}
\end{table*}

We now address our first research question: What are the causal effects of brevity constraints on message success?

Recall that our experimental strategy consists of two steps: first, in the content production phase, we extract shortened versions of original tweets, and second, in the content consumption phase, we measure the quality of these shortened versions in comparison to the original, unshortened version. Part of the content production phase is to ensure that shortened versions contain the essential content of the unshortened version. Thus, since length is the only difference between the original, unshortened tweets and the tweets shortened to prespecified lengths, systematic differences in quality can be causally attributed to shortening.

One potential confound remains: the original tweets are unedited, whereas all the shortened tweets are edited (in the process of being shortened). Thus, as discussed in \Secref{sec:experiment}, to distinguish the benefits of brevity from the benefits of mere editing, we introduce a baseline where the length is decreased by at least one and at most five characters. This baseline is edited, but not significantly shortened. We find that this slight editing often fixes typos (as illustrated in \Tabref{tab:1}) and causes the baseline to be better than the original tweets.
In total, across all individual comparisons between original tweets and baseline versions, workers preferred the baseline version in 58\% of cases.
When first taking a majority vote over all workers who voted on the same original\slash baseline pair, the baseline was preferred for 65\% of all original tweets. If there are additional benefits of shortening beyond editing, they will result in better performance compared to the baseline.

\xhdr{Example}
It is helpful to discuss an example before studying these effects on a more aggregate level. 
\Tabref{tab:2} contains the output of our pipeline (including the shortened versions, the comprehension questions, and estimated probabilities of success) for one particular original tweet. For the colored tweets, the concise version was more successful than the baseline (in this case, 5 out of 8 shortened versions). The best version of this tweet was the one whose length was 80--90\% of the original length. It was chosen over the original 75\% of the time. Interestingly, the version cut down to 30--40\% of the original length is still better than the original, demonstrating that even substantial shortening can sometimes improve the perceived quality.

\xhdr{Measuring the effect of brevity on message success}
We evaluate the effects of shortening in two ways: length-centric and tweet-centric.

\begin{figure}
        \includegraphics[width=.75\textwidth]{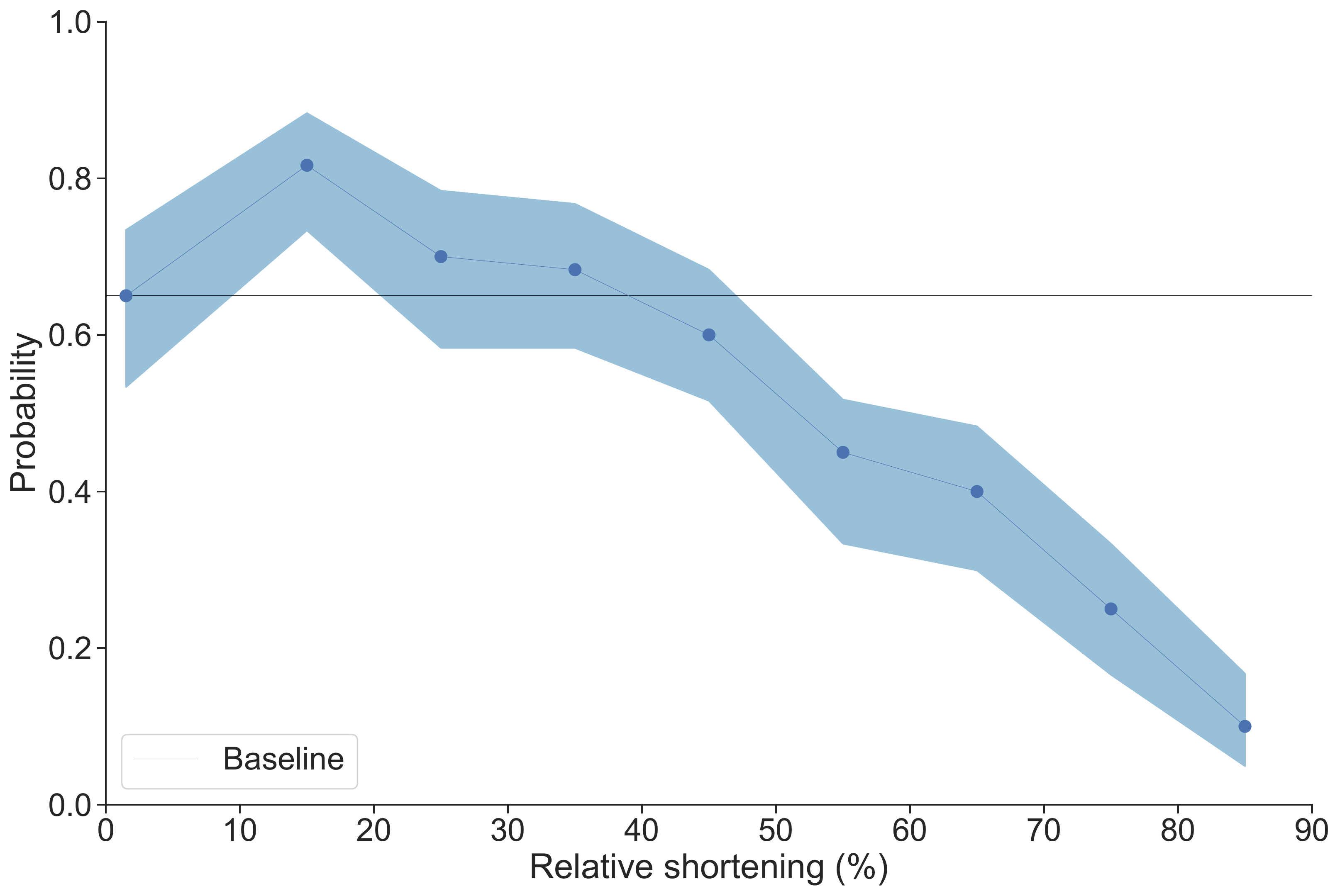}
        \caption{
         Measuring the effect of brevity as a function of the level of shortening (fraction of characters deleted, out of the original 250 characters). Length-centric analysis: fraction of tweets for which the respective shortened version won, according to a majority vote among raters (with bootstrapped 95\% confidence intervals).
         We see that shortening helps: shortened versions beat the baseline for levels of shortening up to 30--40\% of the original length. The optimal range corresponds to shortening by 10--20\% of the original length.
         }
         \label{fig:1}
\end{figure}

\textit{(a) Length-centric analysis.} 
First, we conduct a length-centric analysis. Each bucket of shortening corresponds to the fraction of characters deleted, out of the original 250. Each bucket contains 60 shortened tweets, one for each original tweet. For each tweet, we take a majority vote over the 50 workers to decide which version won---edited or original---and calculate the fraction of tweets in the bucket for which the shortened version won. The results of this analysis are shown in \Figref{fig:1}. 

We observe significant and consistent benefits of shortening. The shortened versions beat the baseline on average (which itself beats the original 65\% of the time) for reductions as high as 30--40\%. Thus, on average, our original 250-character tweets can be reduced to about 165 characters by non-experts and improve in quality as judged by crowd workers. Additionally, edited versions have a higher probability of winning against the original until the length is reduced to up to 30--40\% of the original length. Finally, we observe that the optimal relative shortening is by 10--20\% of the original length. 

\begin{figure}
        \includegraphics[width=.75\textwidth]{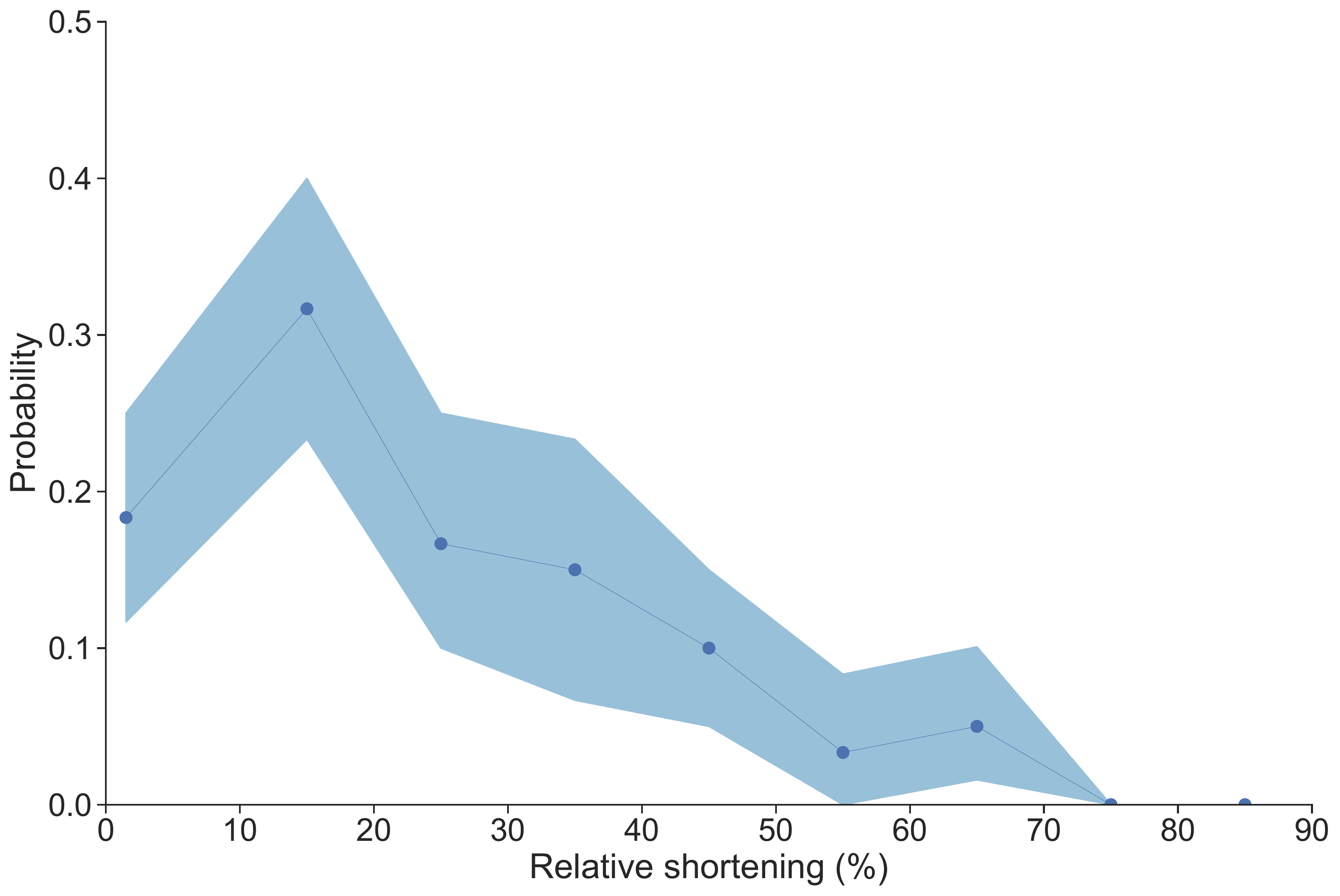}
        \caption{
         Measuring the effect of brevity as a function of the level of shortening (fraction of characters deleted, out of the original 250 characters). Tweet-centric analysis: histogram of the best relative levels of shortening; \ie, for each level of shortening, the fraction of input tweets for which that level saw the largest probability of success (with bootstrapped 95\% confidence intervals). The optimal range corresponds to shortening by 10--20\% of the original length.
         }
         \label{fig:2}
\end{figure}

\textit{(b) Tweet-centric analysis.} 
Second, we conduct a tweet-centric analysis. In our experimental setup, for each original tweet we make 9 comparisons: we compare 8 concise versions and the baseline edited version against the original tweet. At what length are tweets the best? To answer this question, we evaluate for each tweet which length constraint achieved the highest vote fraction compared to the original, and aggregate over all 60 tweets.
This results in the histogram of best per-tweet lengths shown in \Figref{fig:2}. 
As with the previous analysis, the best range is when the original length is reduced by 10--20\%.

We also report our observation that, strikingly, it is never the case that the original tweet is best; for each tweet, there is some edited version that received more than 50\% of votes when compared to the original.
In the median, three edited versions score better against the original than the baseline does.

\begin{figure*}
    \begin{minipage}[t]{.3\textwidth}
        \centering
        \includegraphics[width=\textwidth]{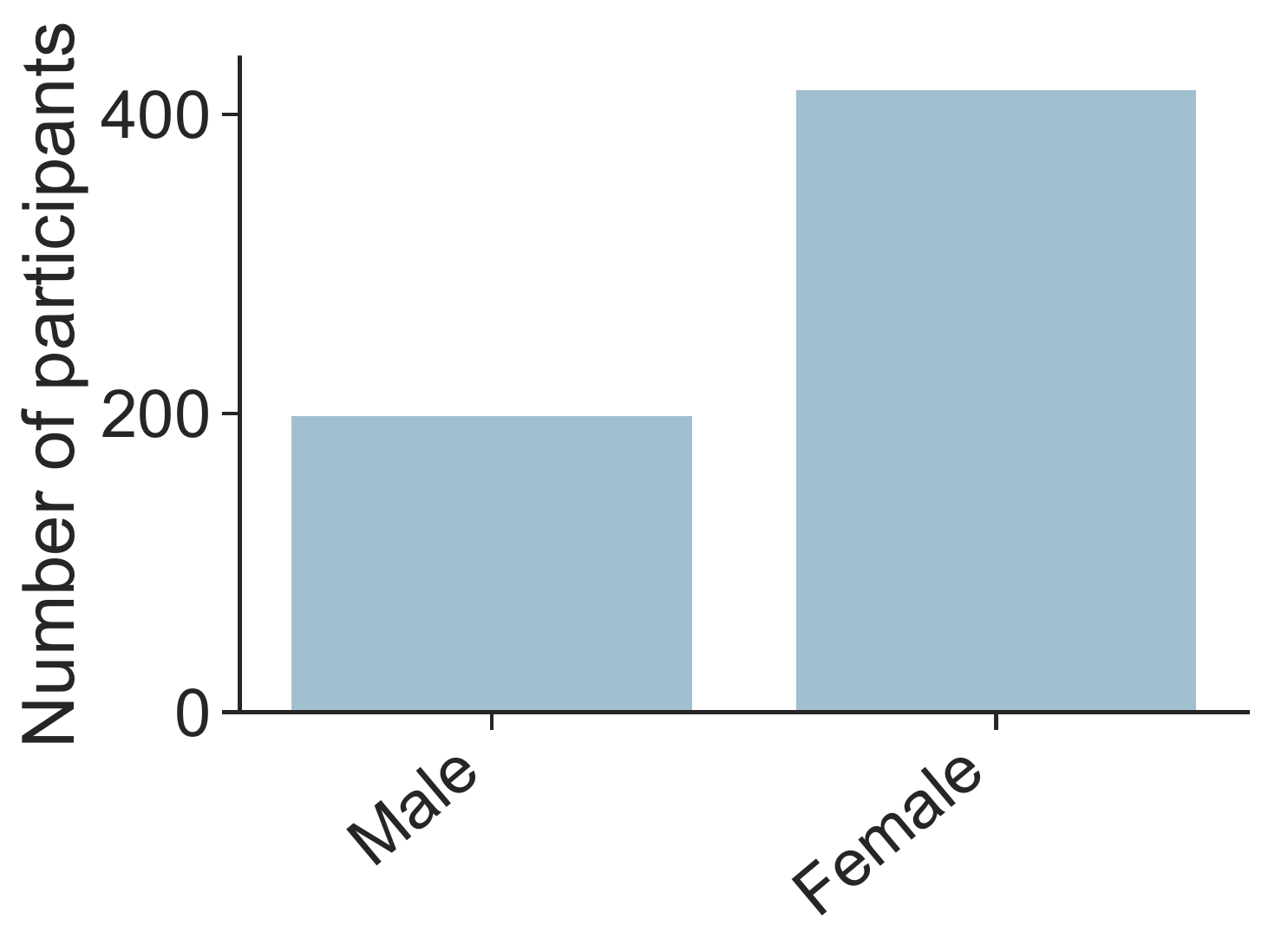}
        \subcaption{Gender}
    \end{minipage}
    \hfill
    \begin{minipage}[t]{.3\textwidth}
        \centering
        \includegraphics[width=\textwidth]{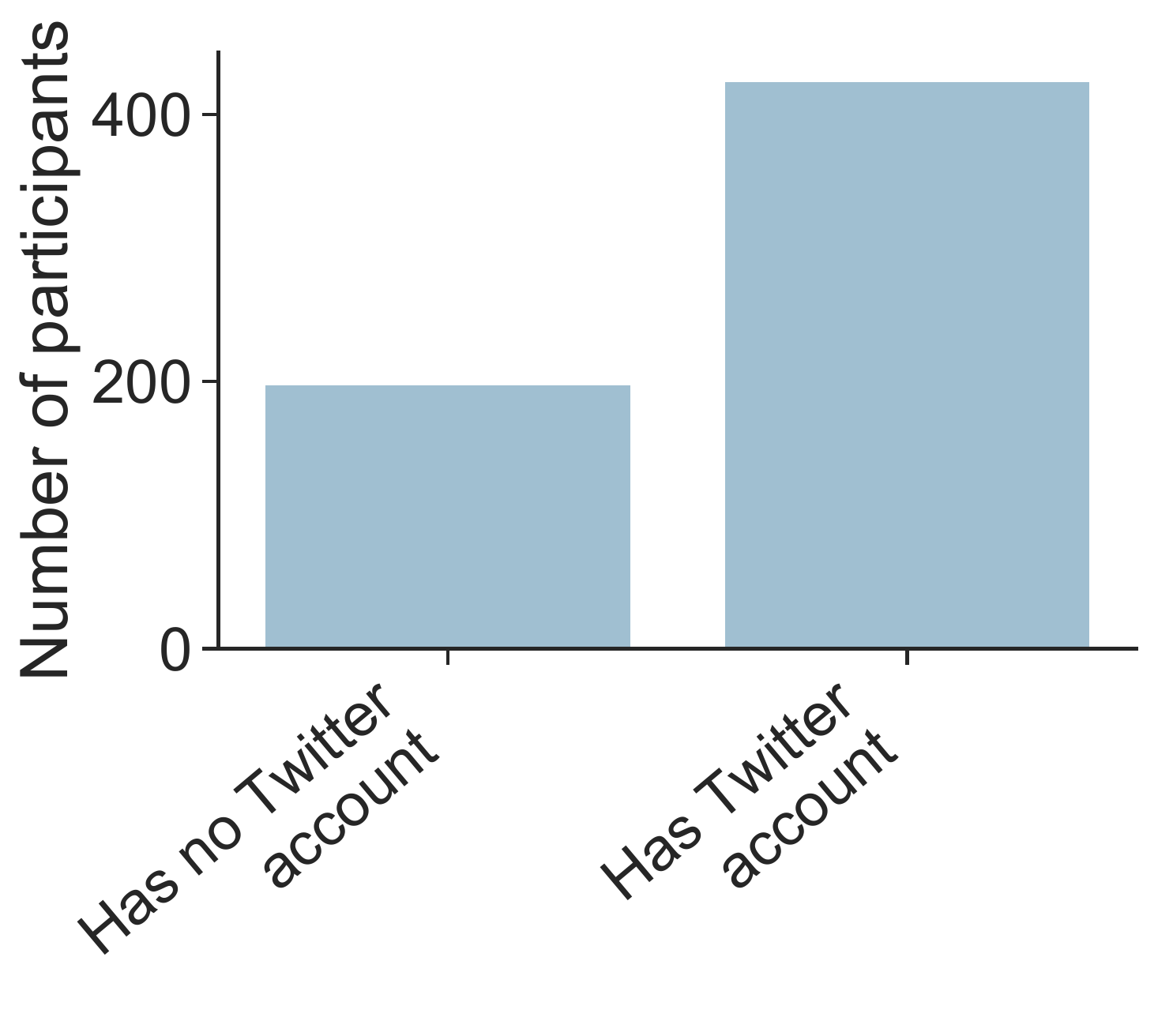}
        \subcaption{Twitter account}
    \end{minipage}
    \hfill
    \begin{minipage}[t]{.3\textwidth}
        \centering
        \includegraphics[width=\textwidth]{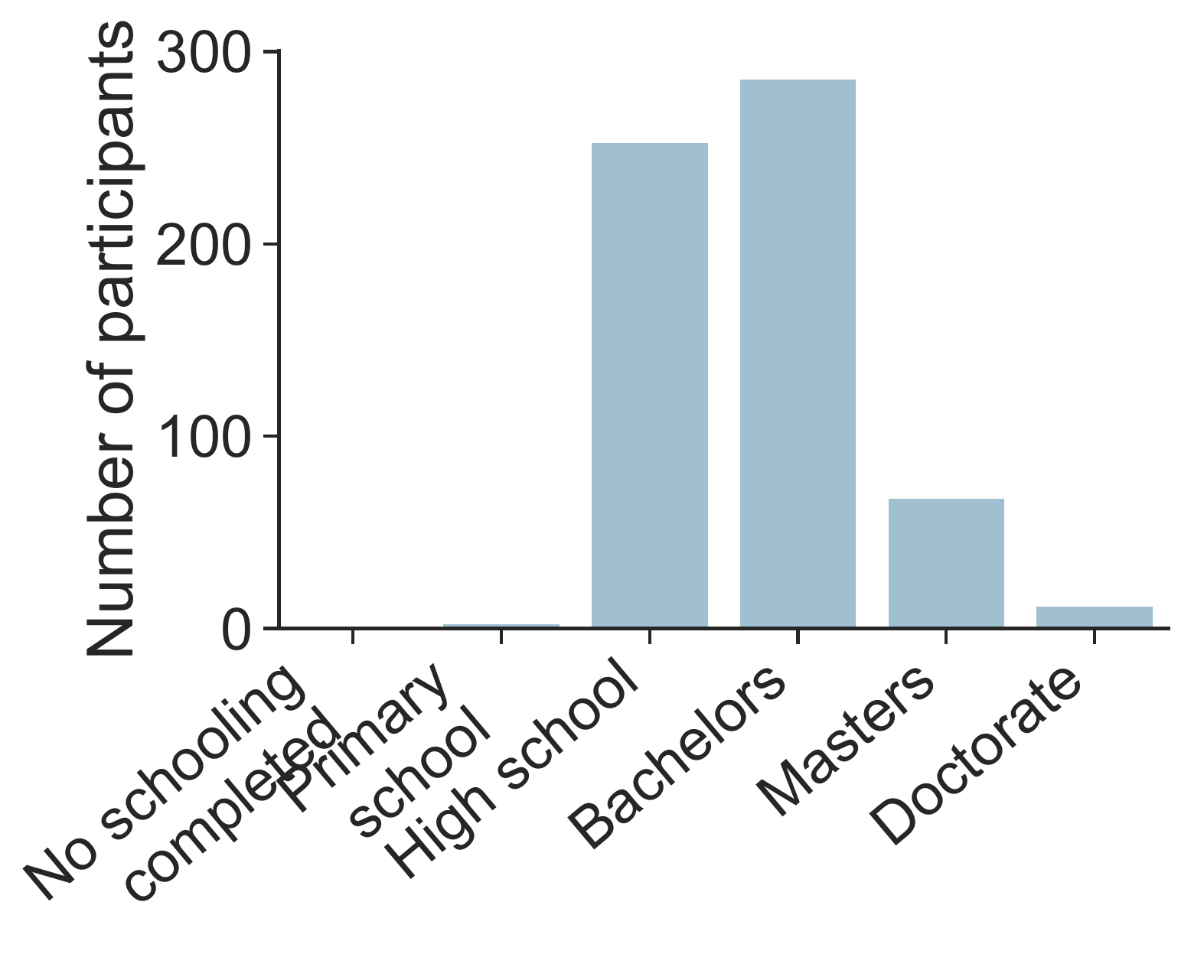}
        \subcaption{Education}
    \end{minipage}
    \hfill
    \begin{minipage}[t]{.3\textwidth}
        \centering
        \includegraphics[width=\textwidth]{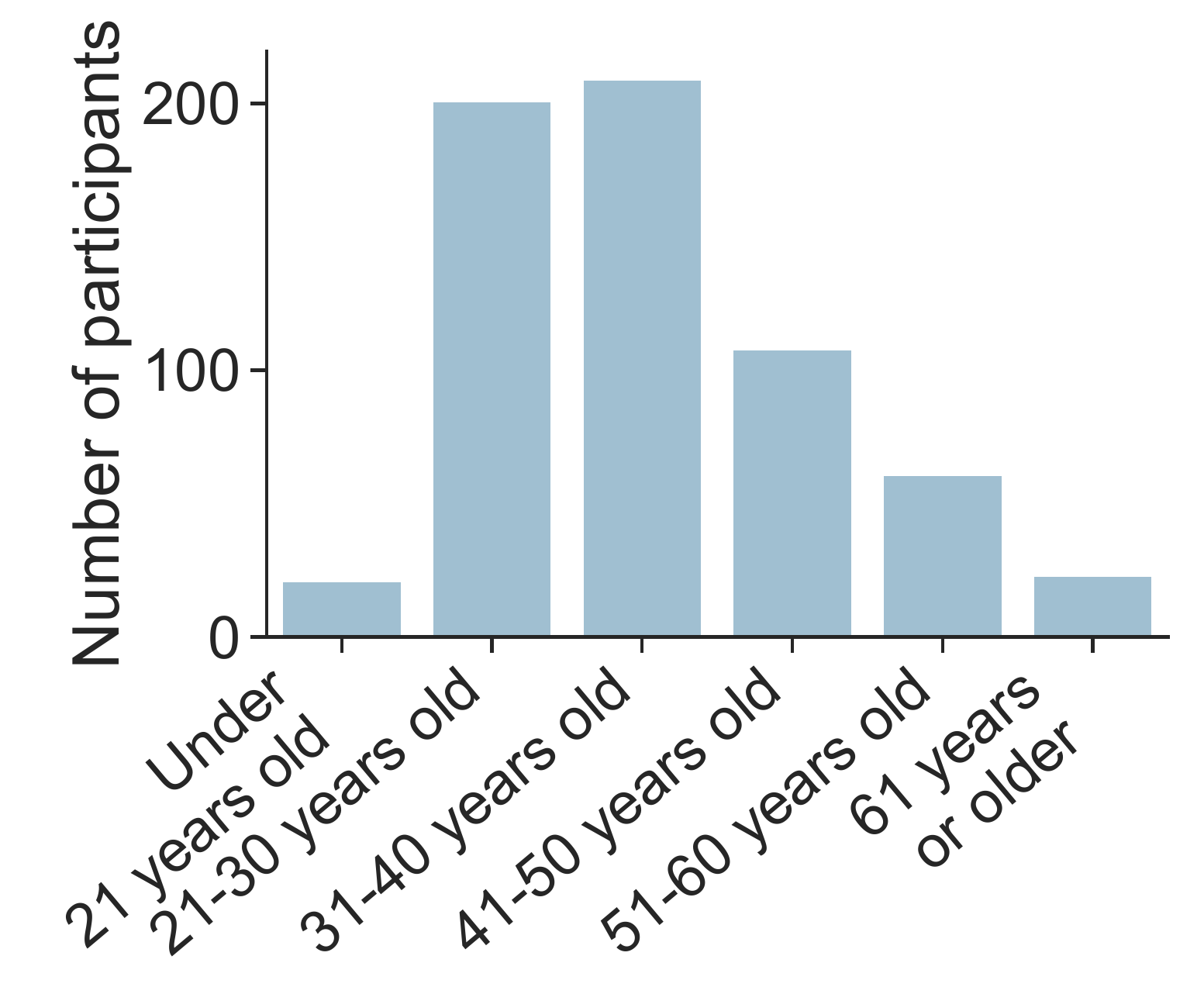}
        \subcaption{Age}
    \end{minipage}  
    \hfill
    \begin{minipage}[t]{.3\textwidth}
        \centering
        \includegraphics[width=\textwidth]{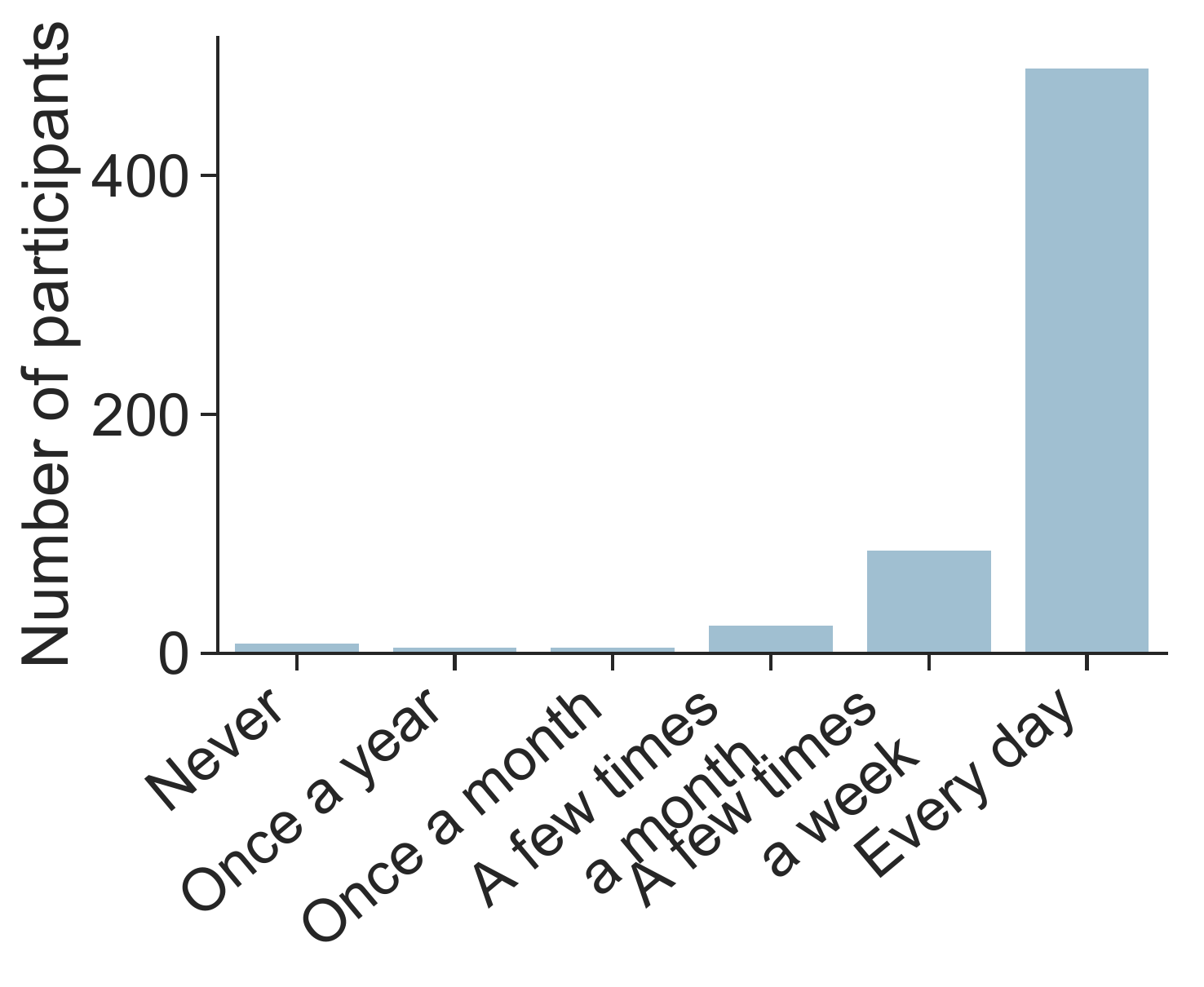}
        \subcaption{Social media usage}
    \end{minipage}
     \hfill
    \begin{minipage}[t]{.3\textwidth}
        \centering
        \includegraphics[width=\textwidth]{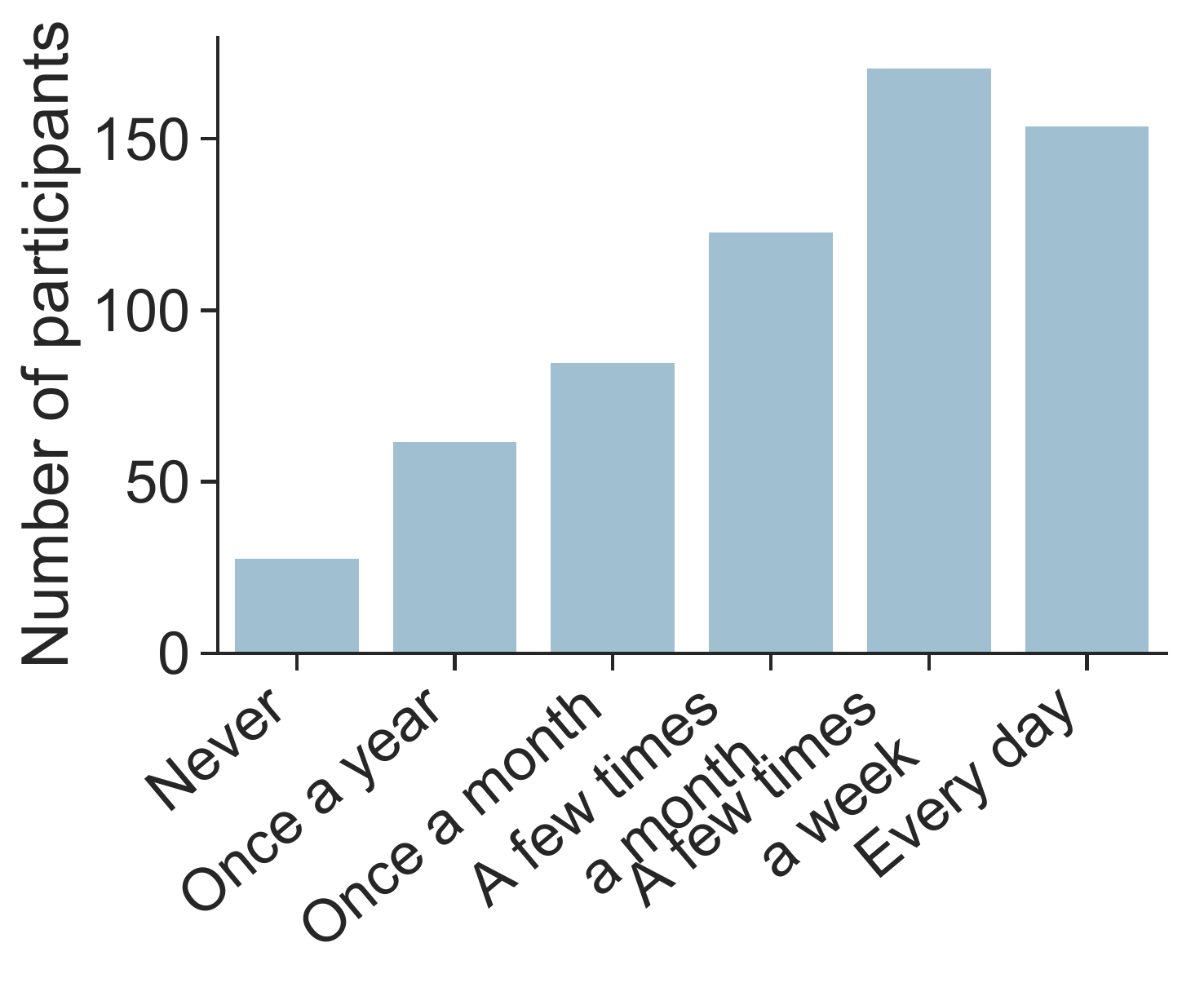}
        \subcaption{Book-reading habits}
    \end{minipage}
    \caption{
    Distributions of worker characteristic on the rating task (Task 5; $N = 623$).}
    \label{fig:demogr}
\end{figure*}

\begin{figure*}[b!]
    \begin{minipage}[t]{.85\textwidth}
        \centering
        \includegraphics[width=\textwidth]{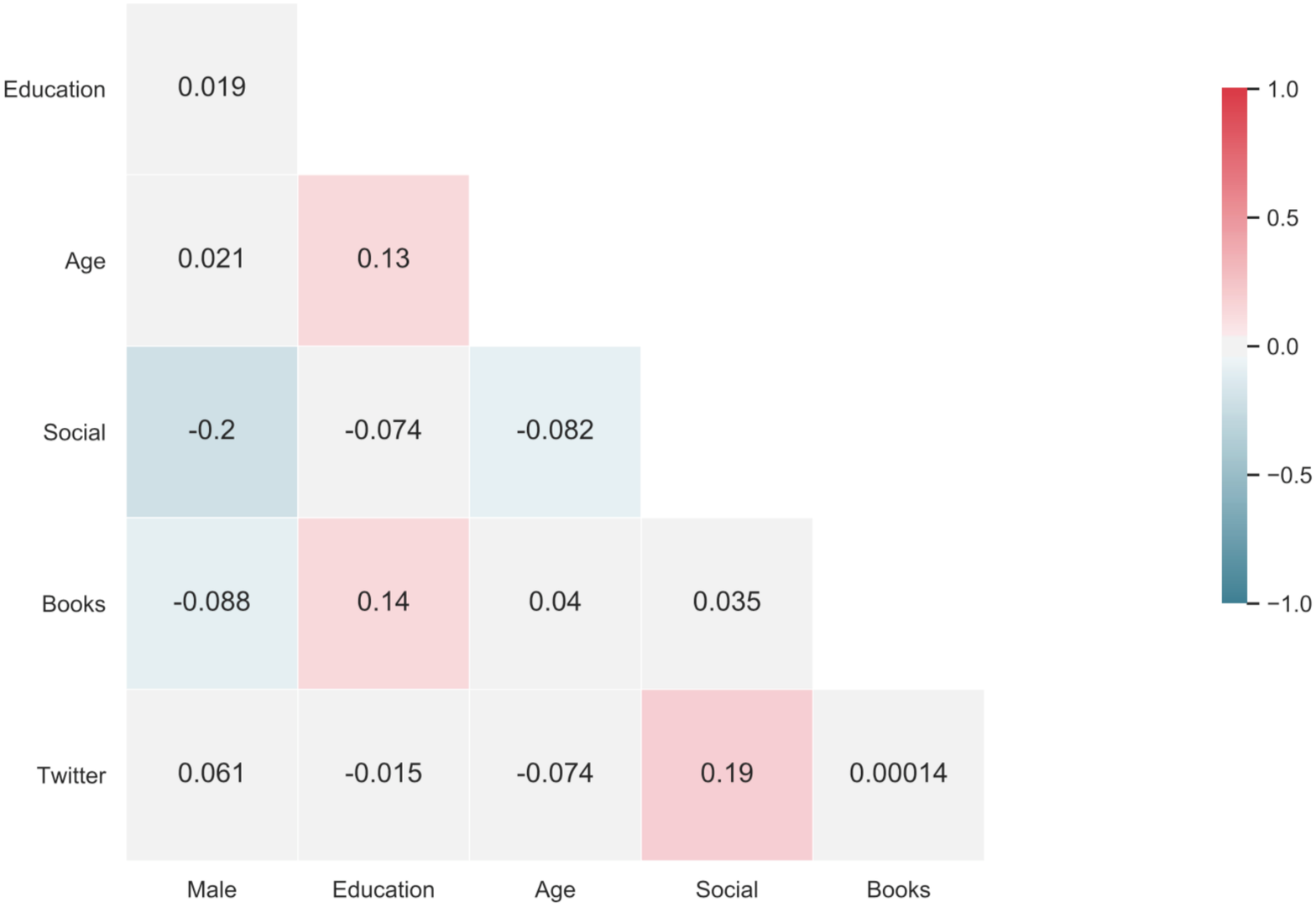}
    \end{minipage}

    \caption{Correlation matrix of participants' demographic, online presence, and reading habits features, introduced in \Figref{fig:demogr}. Nonsignificant correlations are colored gray.}
    \label{fig:corr}
\end{figure*}

\begin{figure*}
    \begin{minipage}[t]{.3\textwidth}
        \centering
        \includegraphics[width=\textwidth]{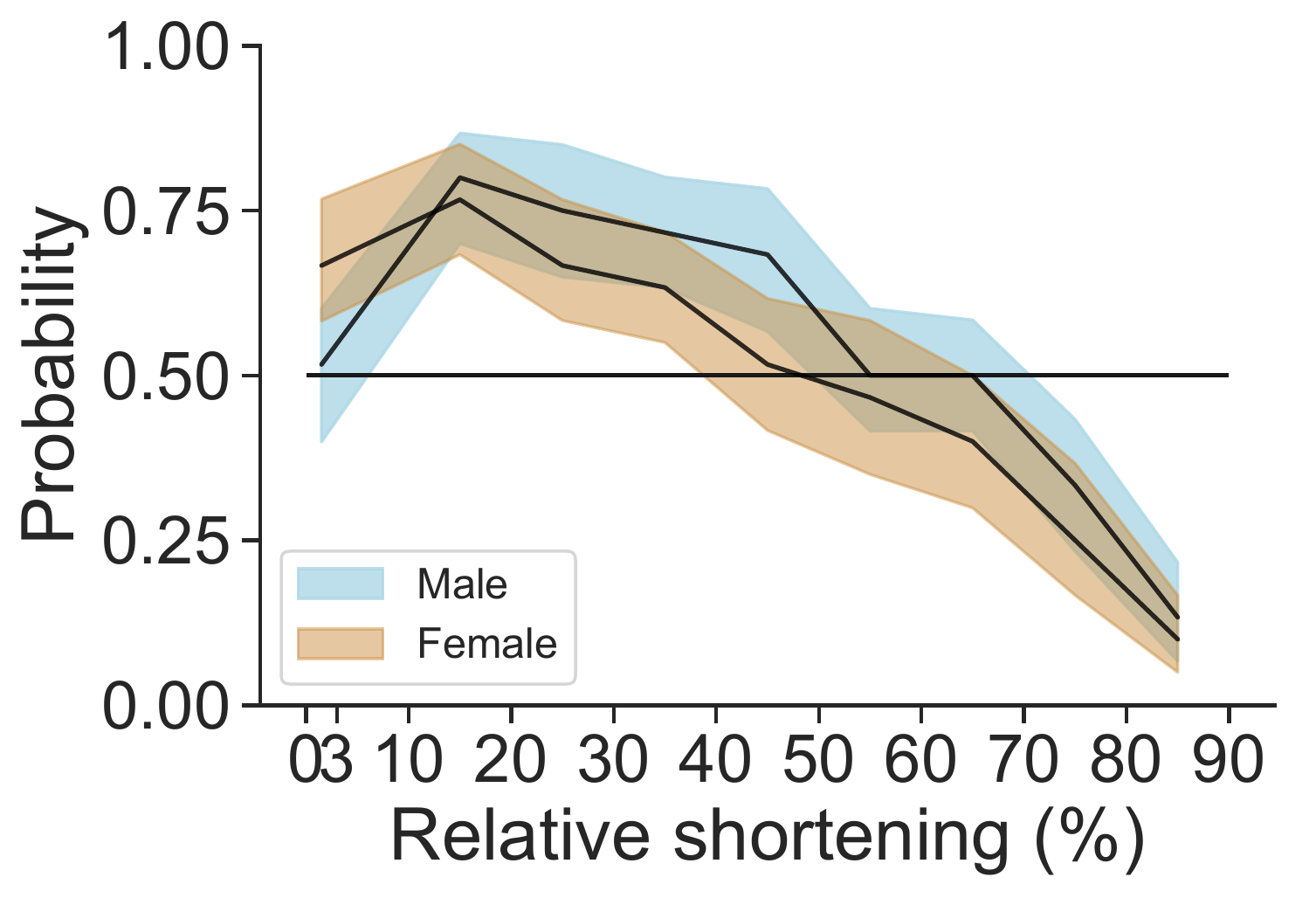}
        \subcaption{Gender}
    \end{minipage}
    \hfill
    \begin{minipage}[t]{.3\textwidth}
        \centering
        \includegraphics[width=\textwidth]{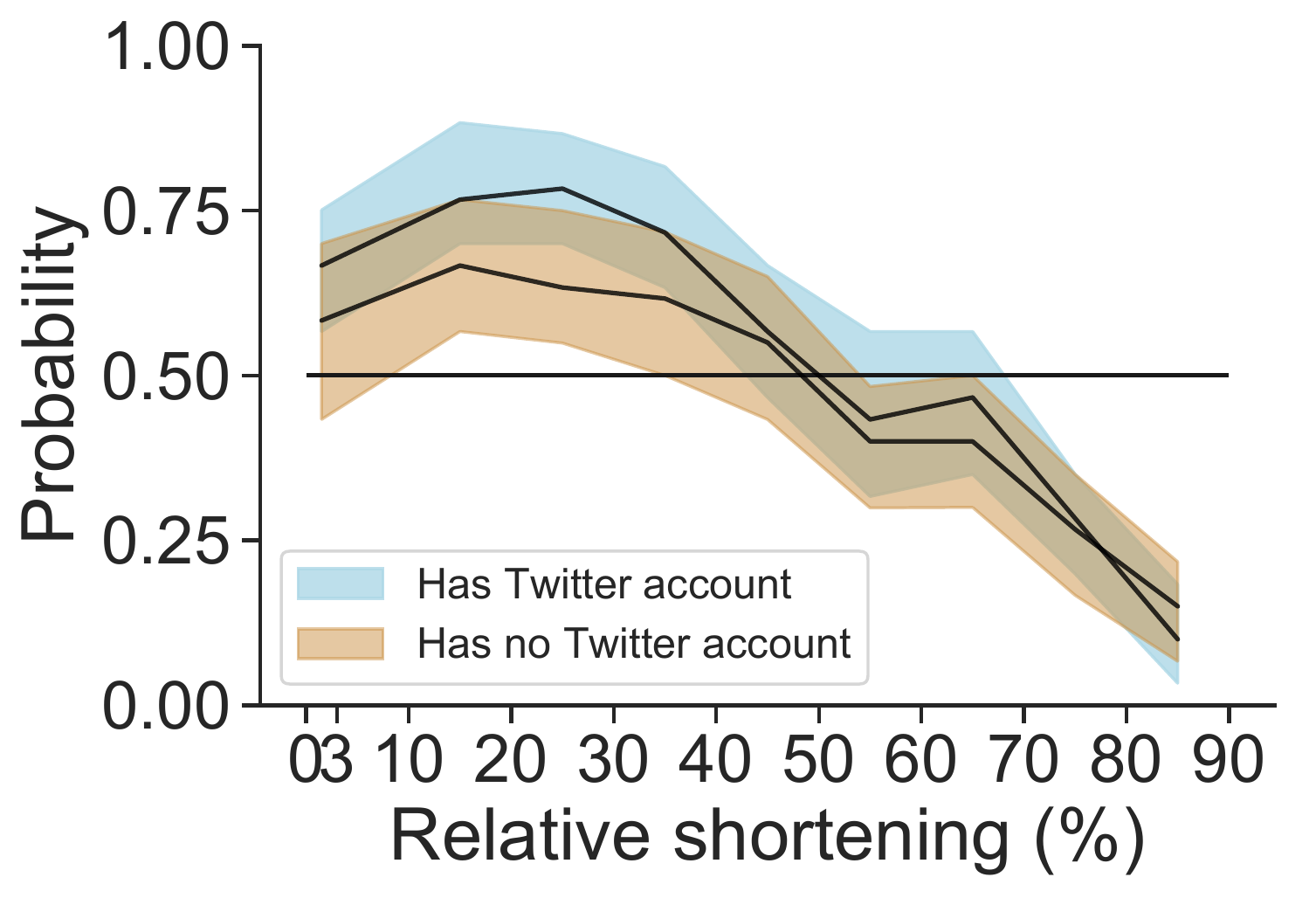}
        \subcaption{Twitter account}
    \end{minipage}
    \hfill
    \begin{minipage}[t]{.3\textwidth}
        \centering
        \includegraphics[width=\textwidth]{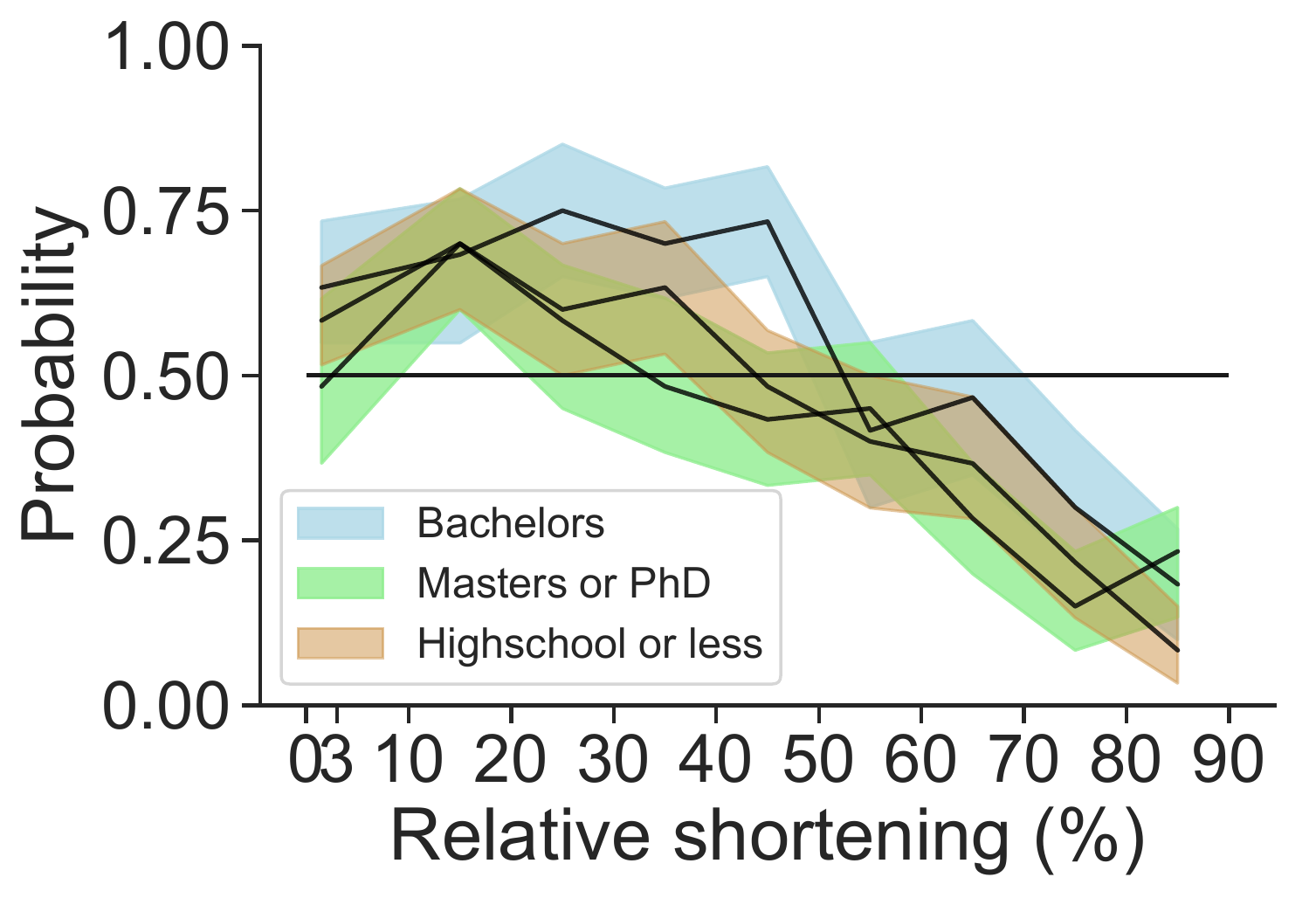}
        \subcaption{Education}
    \end{minipage}
    \hfill
    \begin{minipage}[t]{.3\textwidth}
        \centering
        \includegraphics[width=\textwidth]{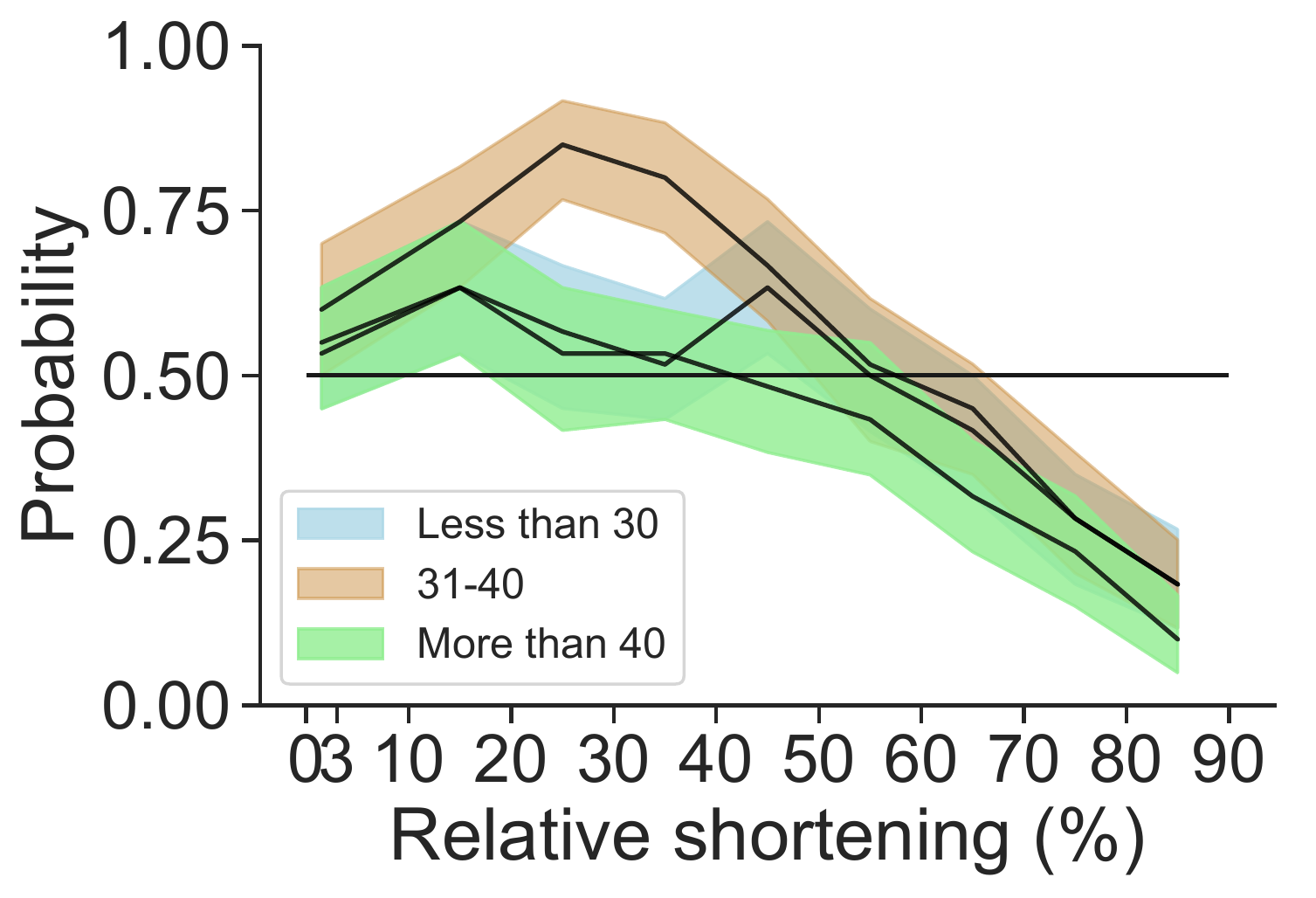}
        \subcaption{Age}
    \end{minipage}  
    \hfill
    \begin{minipage}[t]{.3\textwidth}
        \centering
        \includegraphics[width=\textwidth]{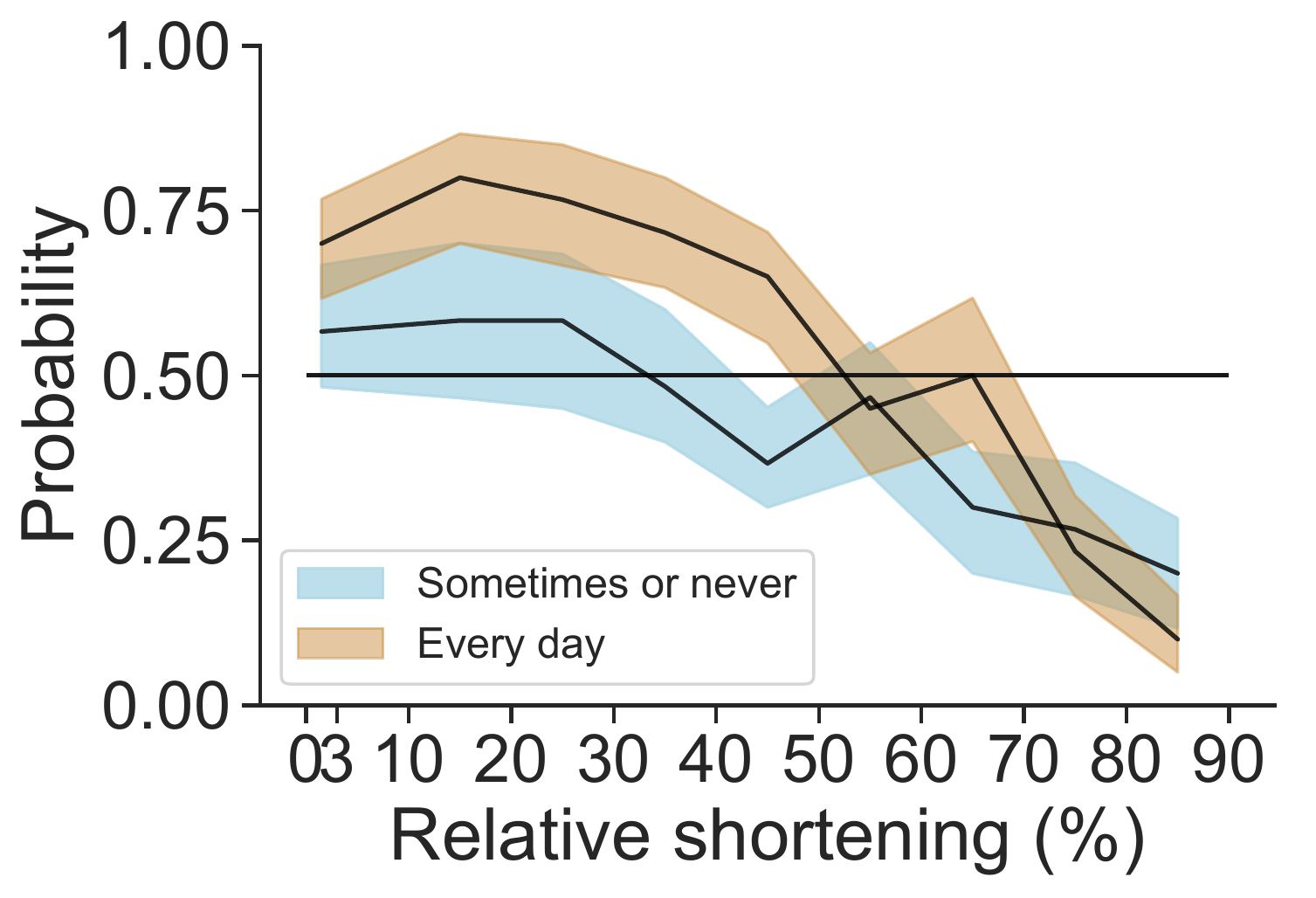}
        \subcaption{Social media usage}
    \end{minipage}
     \hfill
    \begin{minipage}[t]{.3\textwidth}
        \centering
        \includegraphics[width=\textwidth]{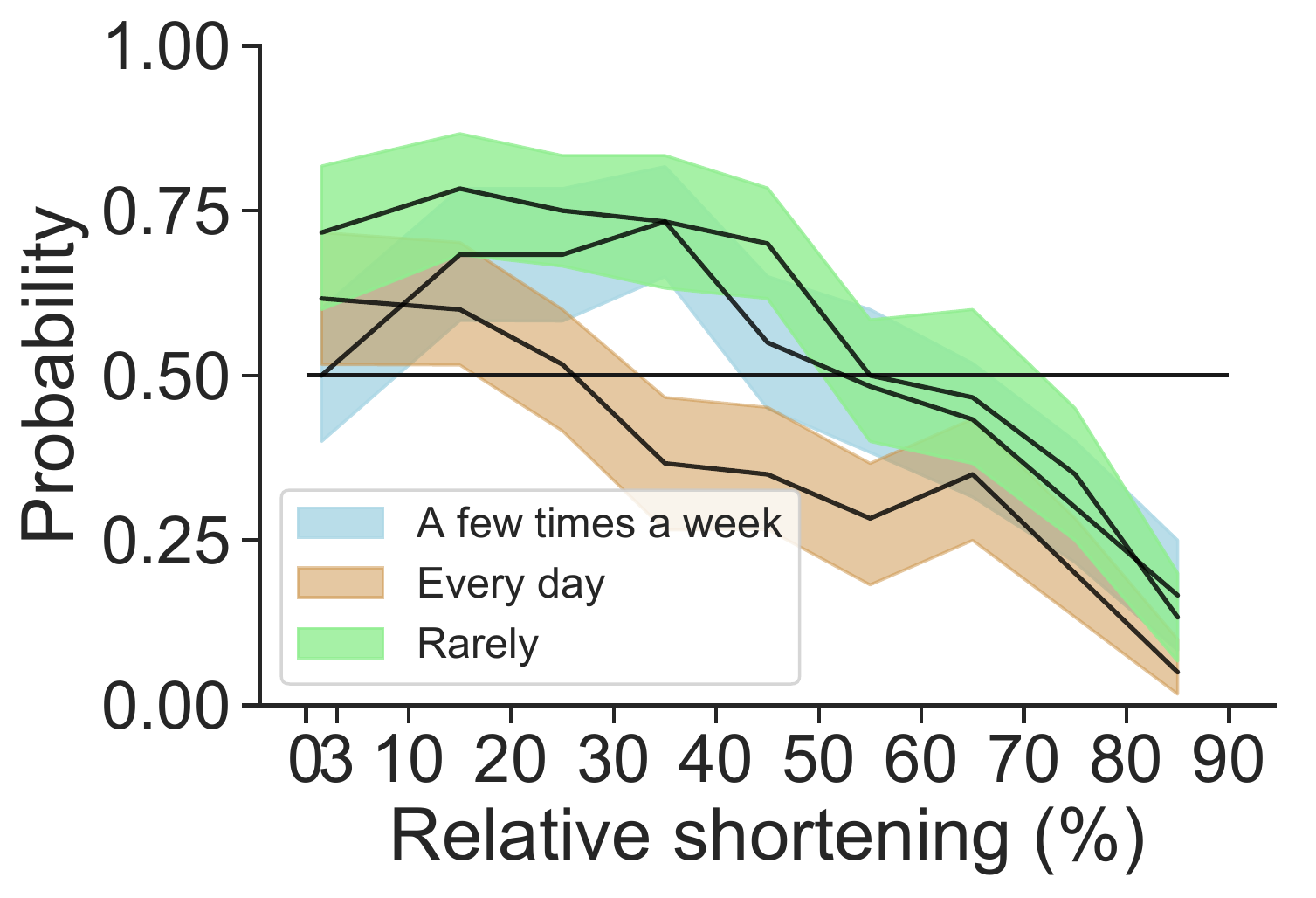}
        \subcaption{Book-reading habits}
    \end{minipage}
    \caption{
    Probability of success based on the majority vote (\cf\ \Figref{fig:1}), conditioned on a given subpopulation of workers, with bootstrapped 95\%  confidence intervals.}
    \label{fig:demogr-succ}
\end{figure*}

\xhdr{Effect of time spent on shortening}
Crowd workers were not limited in the time they could spend on tweet shortening (Task 3). This raises the potential concern that the observed effect of length constraints on the probability of success might simply be explained by spending more time on certain target lengths than on others. To investigate this, we measure how long workers took to shorten tweets for each length constraint. Workers spent on median 48 seconds on the shortening task across all target lengths. Broken down for different target lengths, the median editing time varies from 39 to 55.5 seconds. In the optimal range of shortening (10--20\%), workers spent 53.5 seconds on median, slightly less than in the less successful ranges of 20--30\% and 30--40\% (55.5 seconds on median). Given the significant differences in the probability of success between different target lengths and similar times per task, the differences in time spent alone are unlikely to explain the observed differences in probability of success.

In summary, both a length- and a tweet-centric analysis reveals significant benefits of shortening compared to the baseline, with the optimal amount of shortening being 10--20\% of the original length.

\xhdr{Effects of brevity across subpopulations} Does brevity have the same effect across worker subpopulations? For example, do we observe that younger or older people prefer concise content more?

Although the population of workers in Task~5 is not representative of the general population (\Figref{fig:demogr}), we observe that the results are robust across different subpopulations (\Figref{fig:demogr-succ}).
In \Figref{fig:corr},
we demonstrate how worker characteristics (demographic features, online presence, and reading habits) are correlated. While most of the features are not significantly correlated, we observe some expected correlations, e.g., visiting social media frequently is correlated with having a Twitter account, and the level of education is correlated with age and reading frequency.

To test whether brevity has a significantly different impact on different subpopulations, we use Chow's test \cite{dougherty2011introduction} to test the null hypothesis that the true coefficients (slope and intercept) in two linear regressions are the same across different subpopulations. We apply the test for eight levels of shortening, excluding the baseline. Since we test for six different features, to avoid false positives, we reduce the significance level using Dunn--\v{S}id\'ak correction \cite{vsidak1967rectangular}. The adjusted significance threshold is calculated as $\alpha = 1 - (1-\alpha')^{1/m}$, where $m = 6$ is the number of tests, and  $\alpha' = 0.05$ is the unadjusted significance level, resulting in a lowered significance level $\alpha = 0.0085$.

We observe that people who visit social media every day prefer concise tweets the most ($p<10^{-6}$).
Being active on social media has a stronger influence than having a Twitter account, an insight that hints at the generalizability of the observed effect beyond Twitter. People who read books every day prefer concise tweets the least compared to users with other reading frequencies ($p<010^{-9}$). This observation, together with the fact that reading habits are slightly correlated with a higher level of education (\Figref{fig:corr}), implies that people who read often are more comfortable digesting more complex, verbose content.

\begin{figure}
    \begin{minipage}[]{.95\textwidth}
        \centering
        \includegraphics[width=\textwidth]{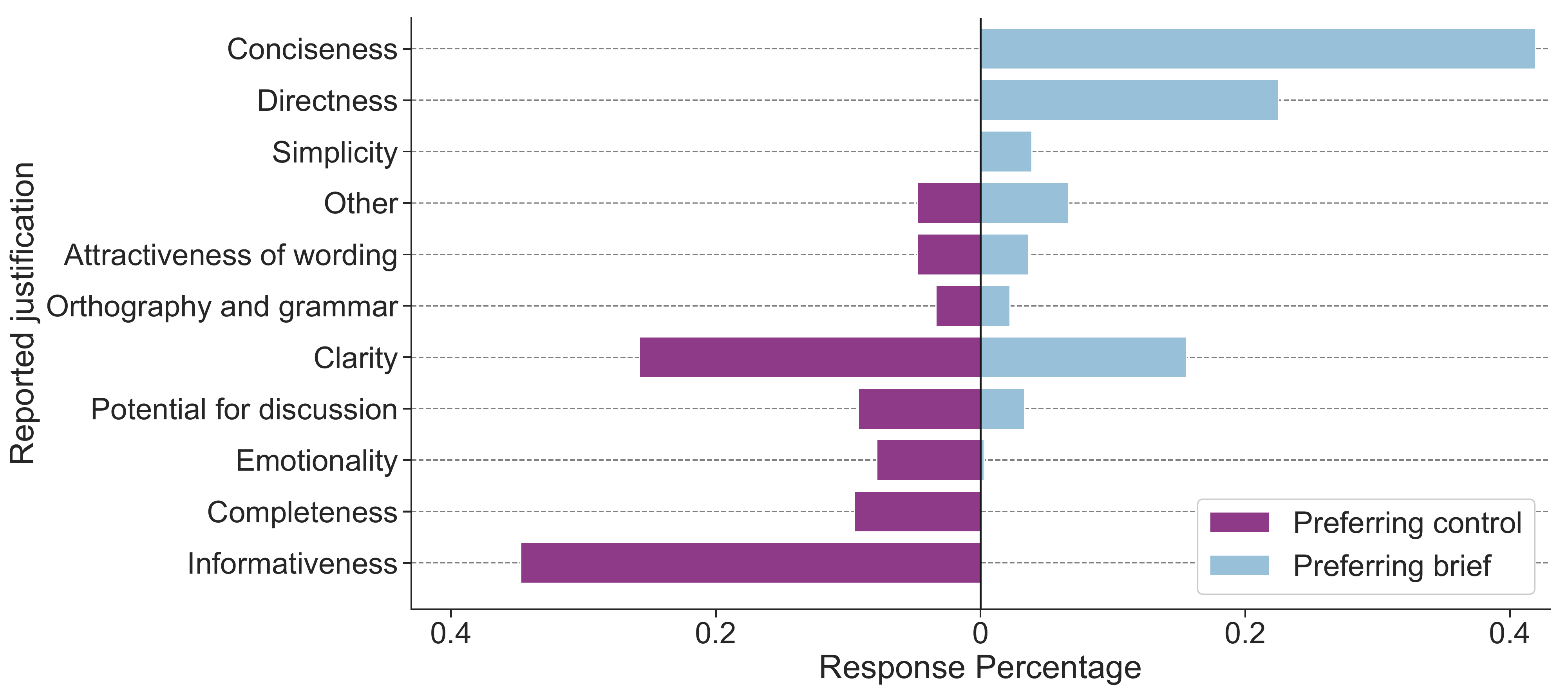}
    \end{minipage}
    \caption{Histograms of reported justifications when brief (blue, right) and control (purple, left) tweets are preferred by workers, sorted by the difference in response percentage, relative to the overall response percentage. When preferring brief tweets, workers mention conciseness, directness and clarity as the top reasons. When preferring original tweets, they mentioned informativeness, completeness, clarity, emotionality, and potential for discussion.}
    \label{fig:reponses}
\end{figure}

\xhdr{Does success imply brevity?}
The relationship between brevity and success could conceivably also be framed the other way around: what if, instead of asking workers to shorten tweets and evaluating them on their perceived success, we asked workers to try to improve tweets (so that they're more likely to be retweeted) and evaluated them on their length? To answer this question, we instructed distinct crowd workers to improve a small subset of the original input tweets with the goal of making the tweet more likely to be retweeted. We substitute the instructions in Task 2 with \textit{Below, you are given a tweet. Your task is to improve the text of the tweet. The goal is to make the tweet more retweeted. The meaning of the original tweet must be maintained. It is important that you don't alter the crucial message by editing the tweet!} We observe that with these instructions, the workers chose to shorten the tweets in most cases, by a median of 16\%. 

\xhdr{Qualitative results}
Before running the large-scale experiment we report in this paper, we conducted a smaller pilot study in which 99 workers who performed the tweet judgment task (Task 5) were also asked to provide free-form justifications for their choices (586 pairwise comparisons in total). Based on these responses, we designed a taxonomy of ten major justifications to characterize the motivations governing why judges chose the way they did in the pairwise comparison.

The responses were then manually coded by one of the authors to assign a label to each response. A single response could mention multiple justifications. The taxonomy was developed by taking a random subsample (one third) of the responses and performing the coding without any initial definitions, resulting in 12 classes. In the next two iterations (each using another third of the data), the taxonomy was consolidated into 10 classes. The 10 classes cover 93\% of the responses. The remaining 7\% are grouped under ``Other''.

We studied the prevalence of the justifications separately for instances where the brief \vs\ the control tweets were preferred. The results are summarized as histograms in \Figref{fig:reponses}. When preferring brief tweets, workers specifically mentioned conciseness, directness (\eg, being on point), and clarity as the top reasons. When preferring original tweets, they mentioned informativeness (\eg, providing more context and more explanations), completeness, clarity, emotionality, and potential for discussion as the top reasons. 

In the final, large-scale experiment, we did not require justifications, but participants could leave optional feedback. This feedback was of high quality and provided similar, clear reasoning, including statements such as ``I feel shorter tweets usually have more impact and cause more discussion as they have a more open ended feel,''
and
``Interesting, different! Sometimes less is more, but not always.''%
\footnote{We note that the feedback was quite positive in general, suggesting that the workers did the task in good faith; \eg,
``Easy to understand and complete. Thank you!''
and
``I truly enjoyed this, and considering this is my primary source of extra income, enjoyment of a task is wonderful.''}

\xhdr{Summary}
In brief, there are significant benefits of brevity. In our experiment, we observe that tweets can be successfully reduced by up to 30--40\% of their original length at no cost in terms of quality. The optimal range of shortening is consistently 10--20\% of the original length.

\section{RQ2: Linguistic traits of brevity}
\label{sec:languagea}

To understand the nature of the observed effect of brevity on perceived quality more deeply, we extend our analysis to include a set of linguistic and psychological features based on Linguistic Inquiry and Word Count (LIWC) \cite{pennebaker2001linguistic}. While we do not study the linguistic aspects in a controlled setup (we only obtain a single edited version for each given length constraint per original tweet), we observe multiple consistent, interesting patterns.

\xhdr{A first look: detecting linguistic aspects typical of concise \vs\ verbose tweets} Our main method of analysis is to compare tokens across different sub-populations of tweets. A token is typical for one set of tweets if it is used frequently within the set, but at the same time unlikely to be used in the other set. Additionally, it is not only the discrepancy between the two probabilities that matters; a word should also appear frequently in a set to be considered typical of the set.
To capture this intuition, we use Jensen--Shannon (JS) divergence, a method of measuring the similarity between two probability distributions. Based on the respective unigram language models, we compute the pointwise JS divergence between the distributions of tokens for original tweets and for concise tweets. In Table \ref{tab:3}, we present the tokens with the largest pointwise JS divergence, separately for tokens with a higher probability of appearing in original \vs\ concise tweets.

Most indicative of original tweets are articles (\textit{the}), linking words (\textit{as, therefore, because, and}), intensifiers (\textit{so}), determiners and quantifiers (\textit{that, any, which}), and certain punctuation marks (\textit{\dots}). These tokens do not carry essential information and frequently are removed in the process of shortening.

Concise tweets, on the other hand, typically contain punctuation that makes it possible to convey the same meaning with a single character: \textit{\&, /,} and {-} are used to substitute longer tokens, such as \textit{and, or,} and \textit{therefore}. Democrats and Republicans are shortened to \textit{dems} and \textit{reps}. Also typical for concise tweets is the usage of punctuation that elicits reactions and conversation, such as \textit{?} and \textit{!}. In the shortening process, workers contract auxiliary verbs to save space without losing any information. As a consequence, contracted forms, such as \textit{'ll} are typical for concise tweets.

Tokens such as \textit{best, say, bad,} and \textit{fail} occur disproportionately in concise tweets as a consequence of editing the original tweets by replacing long, specific phrases with short, more general ones.
For example, the wordy phrasing of the original tweet, ``one of \textbf{the happiest moments} of my life I tend to dwell on sometimes thinking I wish it was still that way [\dots],'' was edited to the simpler and more general phrasing, ``definitely the \textbf{best times} in my life, miss it, and think of it often.'' Similarly, ``\textbf{opinions expressed} in books and movies'' is transformed to ``they \textbf{say} in books and movies.'' 

\begin{table}[b]
  \caption{Detecting linguistic aspects typical for concise \vs\ verbose tweets using pointwise JS divergence.}
  \label{tab:3}
\begin{tabular}{lllll|lllll}
    \toprule
    \multicolumn{5}{l}{\textbf{Top tokens typical for original tweets}} \vline & \multicolumn{5}{l}{\textbf{Top tokens typical for concise tweets}}\\
    \midrule
    that & just & as & ... & and & / & \& & ! & dems & 'll \\
therefore & the & any & so & which &  - & bad & ? & with & \$  \\
has & it  & place & sometimes & because &  best & " & say & reps & fail\\

  \end{tabular}
\end{table}

\begin{figure}
    \begin{minipage}{0.24\columnwidth}
        \centering
        \includegraphics[width=\textwidth]{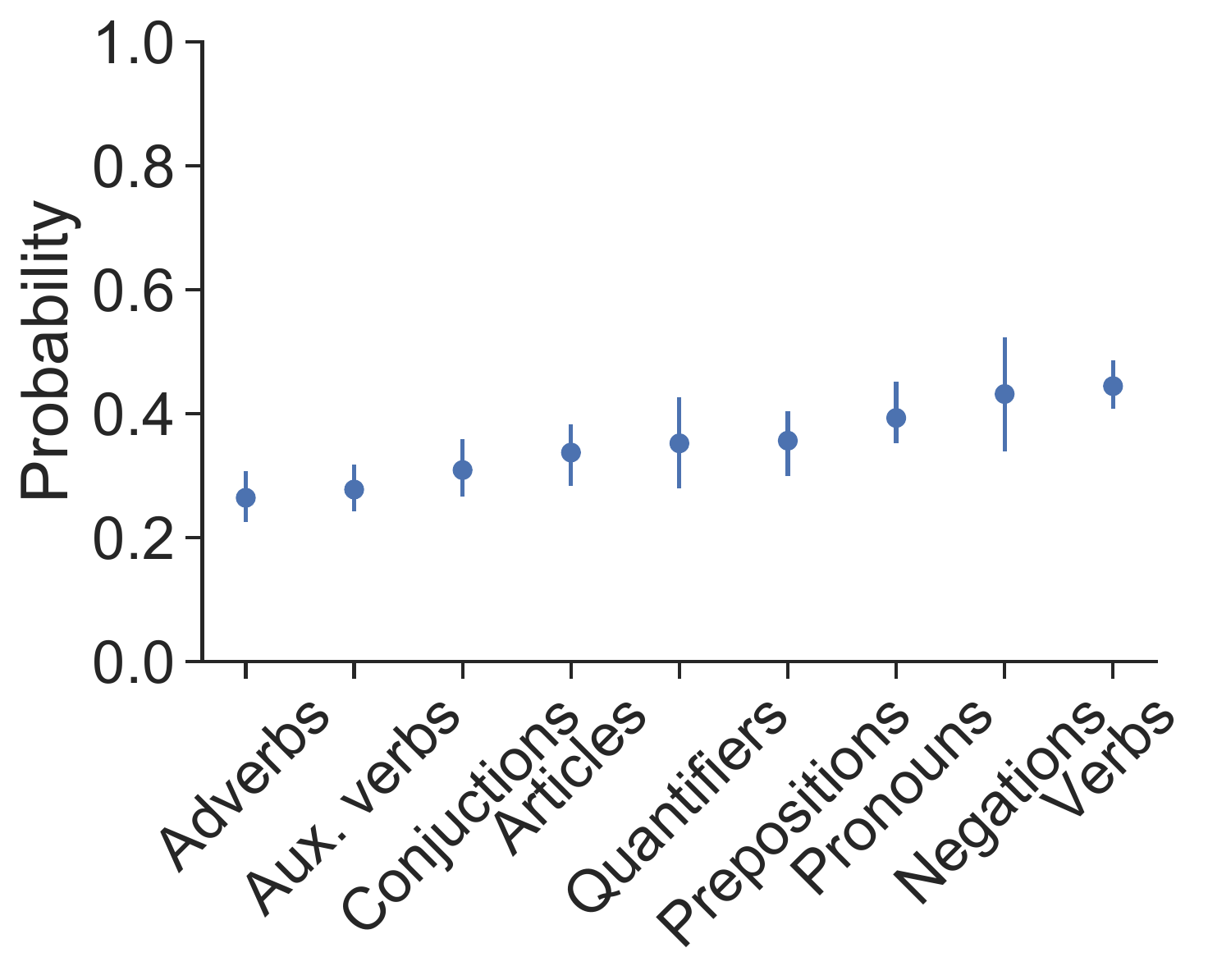}
        \subcaption{All levels of shortening}
        \label{fig:5}
    \end{minipage}
    \hfill
    \begin{minipage}{.24\columnwidth}
        \centering
        \includegraphics[width=\textwidth]{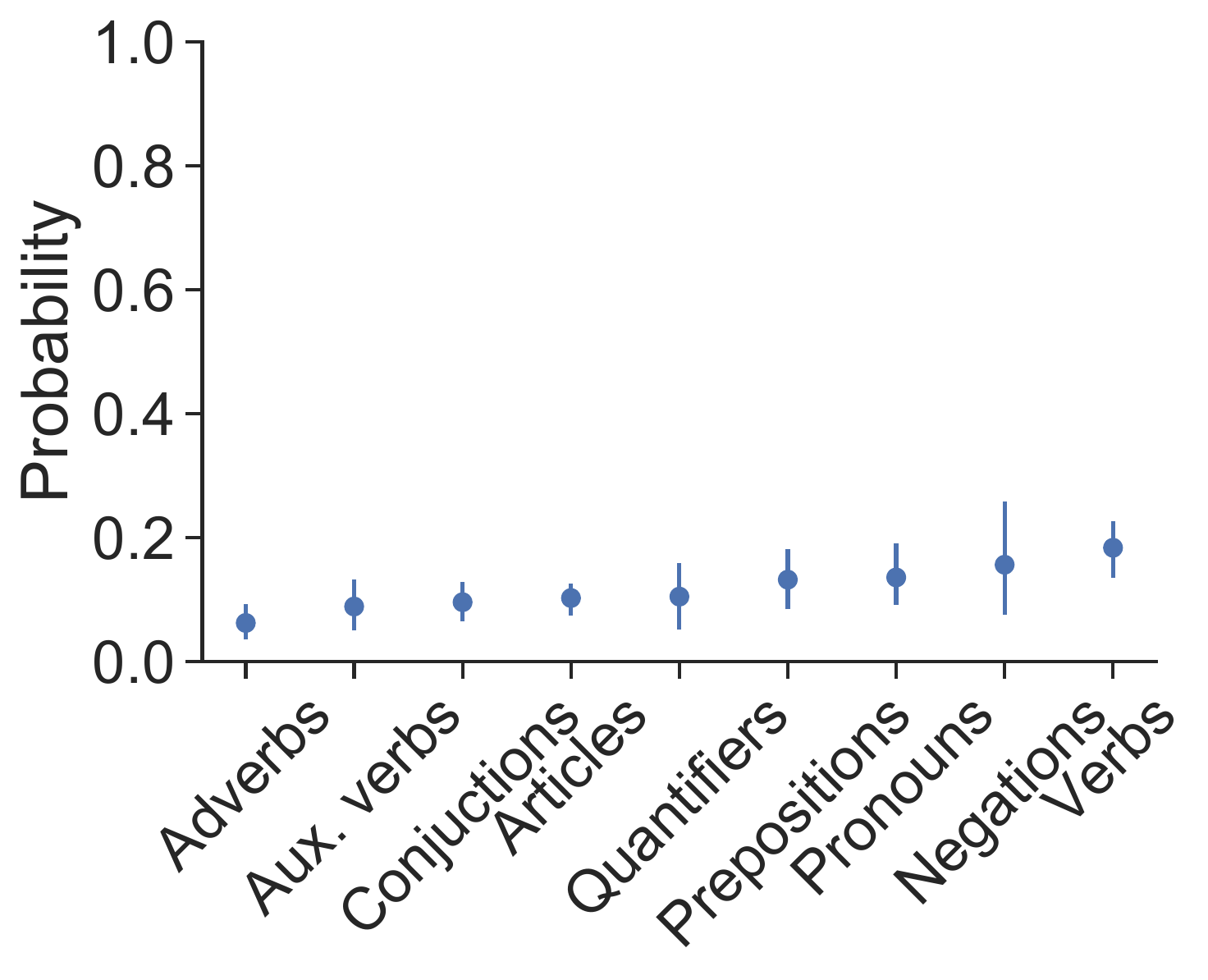}
        \subcaption{Drastic shortening}
        \label{fig:18}
    \end{minipage}
    \hfill
    \begin{minipage}{.24\columnwidth}
        \centering
        \includegraphics[width=\textwidth]{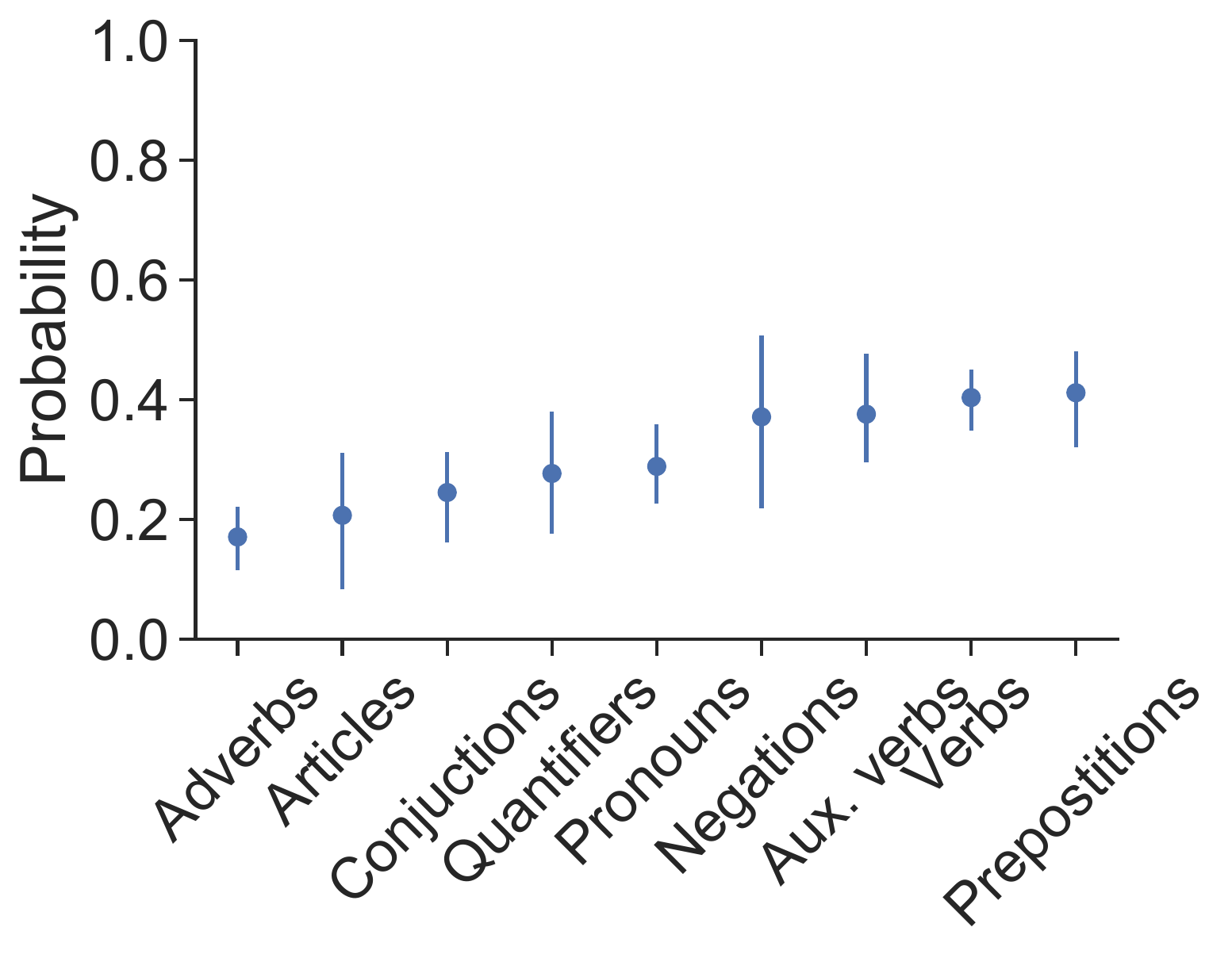}
        \subcaption{Medium shortening}
        \label{fig:19}
    \end{minipage}
    \hfill
    \begin{minipage}{.24\columnwidth}
        \centering
        \includegraphics[width=\textwidth]{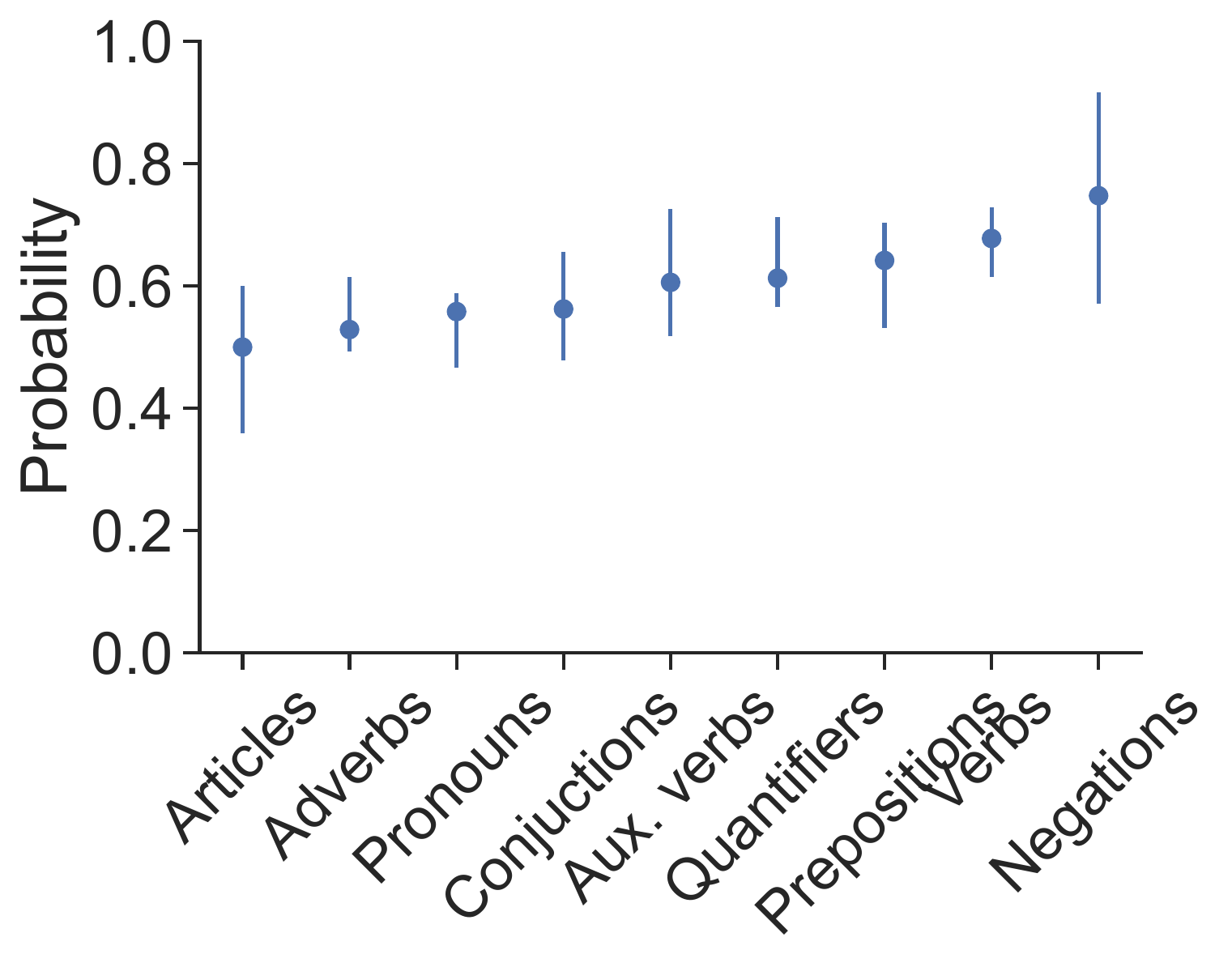}
        \subcaption{Slight shortening}
        \label{fig:20}
    \end{minipage}
    \caption{Preservation analysis:
    probability of preservation in the shortening process (with bootstrapped  95\% confidence intervals) for various parts of speech, based on linguistic process categories from LIWC \cite{pennebaker2001linguistic}.
    Levels of shortening are
    (a) 10--90\%,
    (b) 60--90\%,
    (c) 40--60\%,
    (d) 10--40\%.
    Articles, adverbs, conjunctions and auxiliary verbs have the lowest probability of being preserved. Parts of speech that convey essential information, such as verbs and negations, have the highest probability of being preserved.
    }
    \label{fig:preservation}
\end{figure}

\xhdr{Token-preservation analysis} Following these insights, we continue the analysis in a more  fine-grained way by explicitly tracking each unique token. 
Here, we are interested in comparing every shortened tweet with its corresponding original tweet and tracking each token to tell whether it was added, modified, or deleted. 
For each shortened-original tweet pair, we do this by using standard dynamic programming with insertion, deletion and substitution operations to find the shortest sequence of edit operations for transforming the original into the shortened tweet. 
We define the probability of a token being preserved as the fraction of its occurrences in original tweets that were kept or only slightly modified (character\hyp based edit distance 2 or less). Edit operations are allowed for tokens at least five characters long, and, to avoid falsely detecting edits, the original and the transformed token must have an edit distance of two characters or less. That way we allow for edits such as reformulating by changing the tense, converting plurals to singulars, or introducing or fixing typos. 

The results are visualized in \Figref{fig:preservation}.
Consistent with the previous analysis, articles, adverbs, conjunctions and auxiliary verbs have the lowest probability of being preserved. On the other hand, parts of speech that convey essential information, such as verbs and negations, have the highest probability of being preserved (\Figref{fig:5}). These findings are consistent across different levels of shortening, from minimal to drastic (\Figref{fig:preservation}b--d).

\begin{table}[b]
  \caption{Analysis of subtypes of affect, averaged across all levels of shortening. Tokens carrying negative affect are more likely to be kept, compared to positive emotion. The trend is consistent across subtypes of negative affect (anger, sadness, and anxiety).
  }
  \label{tab:4}
\begin{tabular}{llr}
    \toprule
    \multicolumn{2}{l}{\textbf{Type of affect}} & \textbf{Probability of being preserved} \\
    \midrule
    \multicolumn{2}{l}{Positive emotion} & 0.54\\
    \multicolumn{2}{l}{Negative emotion} & 0.64\\
    & Anger & 0.65\\
    & Sadness & 0.63 \\
    & Anxiety & 0.67\\
  \end{tabular}
\end{table}

\begin{figure}
    \begin{minipage}{.3\textwidth}
        \centering
        \includegraphics[width=\textwidth]{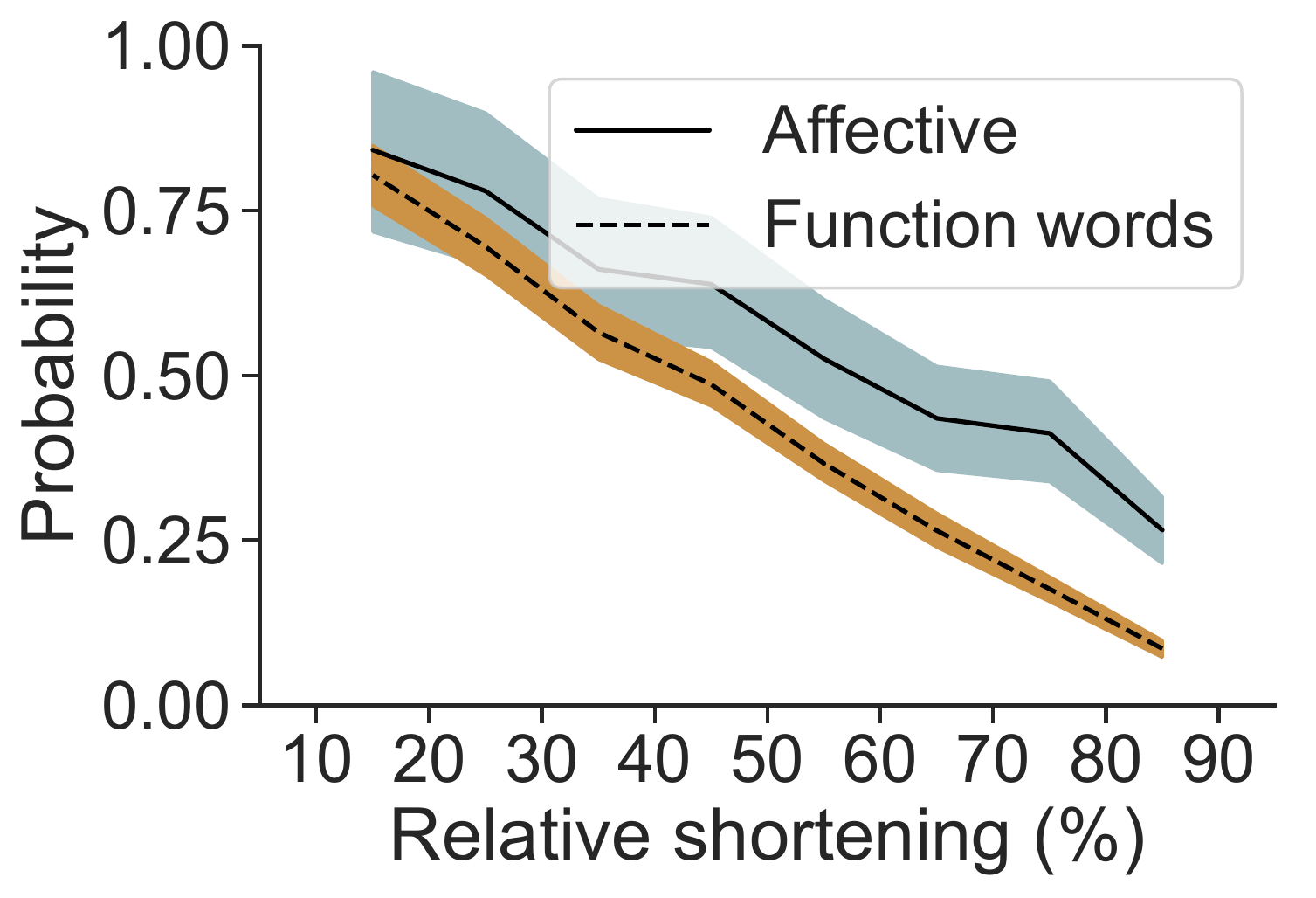}
        \subcaption{Affective*}
        \label{fig:10}
    \end{minipage}
    \hfill
    \begin{minipage}{.3\textwidth}
        \centering
        \includegraphics[width=\textwidth]{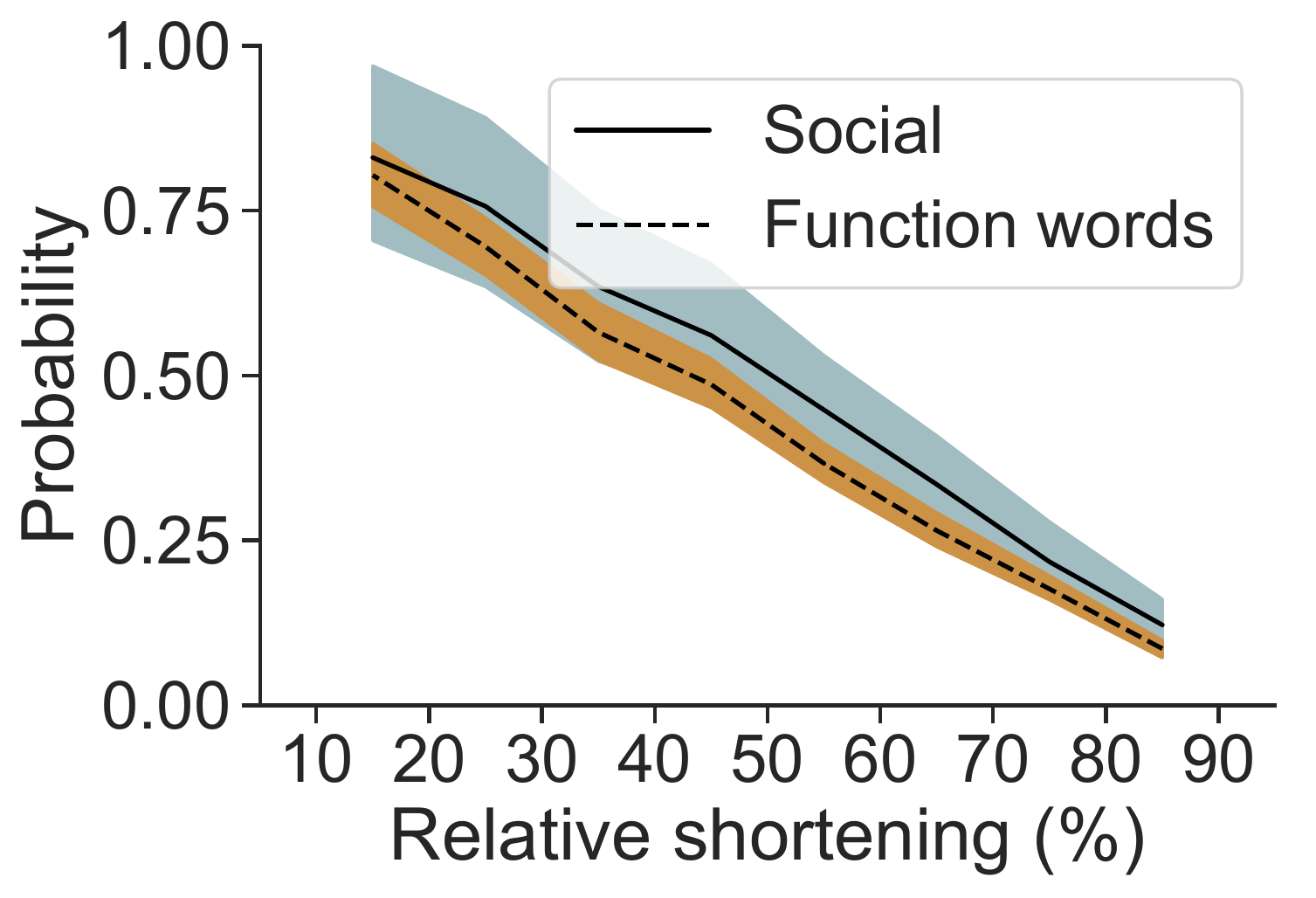}
        \subcaption{Social*}
        \label{fig:11}
    \end{minipage}
        \hfill
    \begin{minipage}{.3\textwidth}
        \centering
        \includegraphics[width=\textwidth]{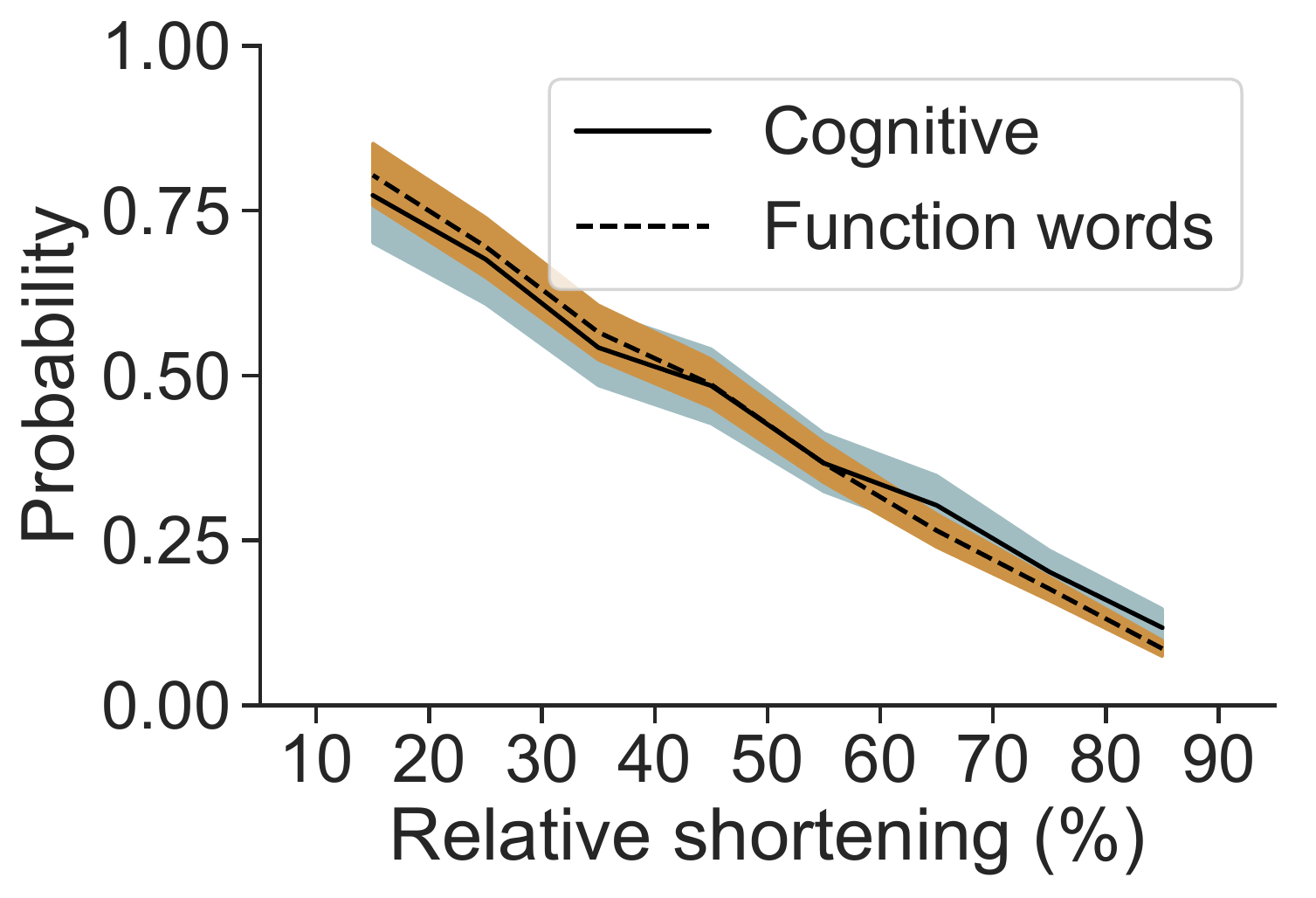}
        \subcaption{Cognitive}
        \label{fig:12}
    \end{minipage}
        \hfill
    \begin{minipage}{.3\textwidth}
        \centering
        \includegraphics[width=\textwidth]{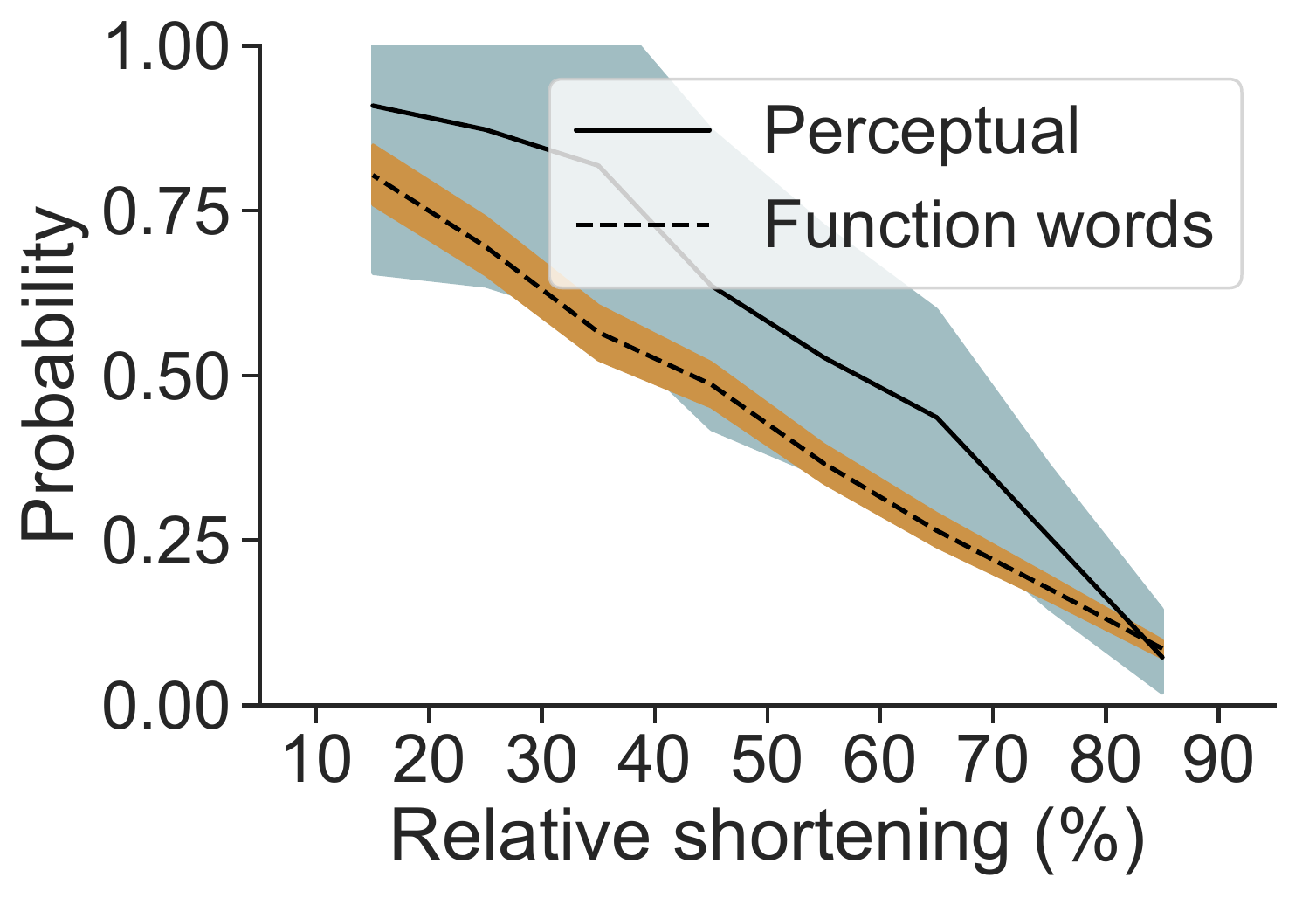}
        \subcaption{Perceptual*}
        \label{fig:13}
    \end{minipage}
        \hfill
    \begin{minipage}{.3\textwidth}
        \centering
        \includegraphics[width=\textwidth]{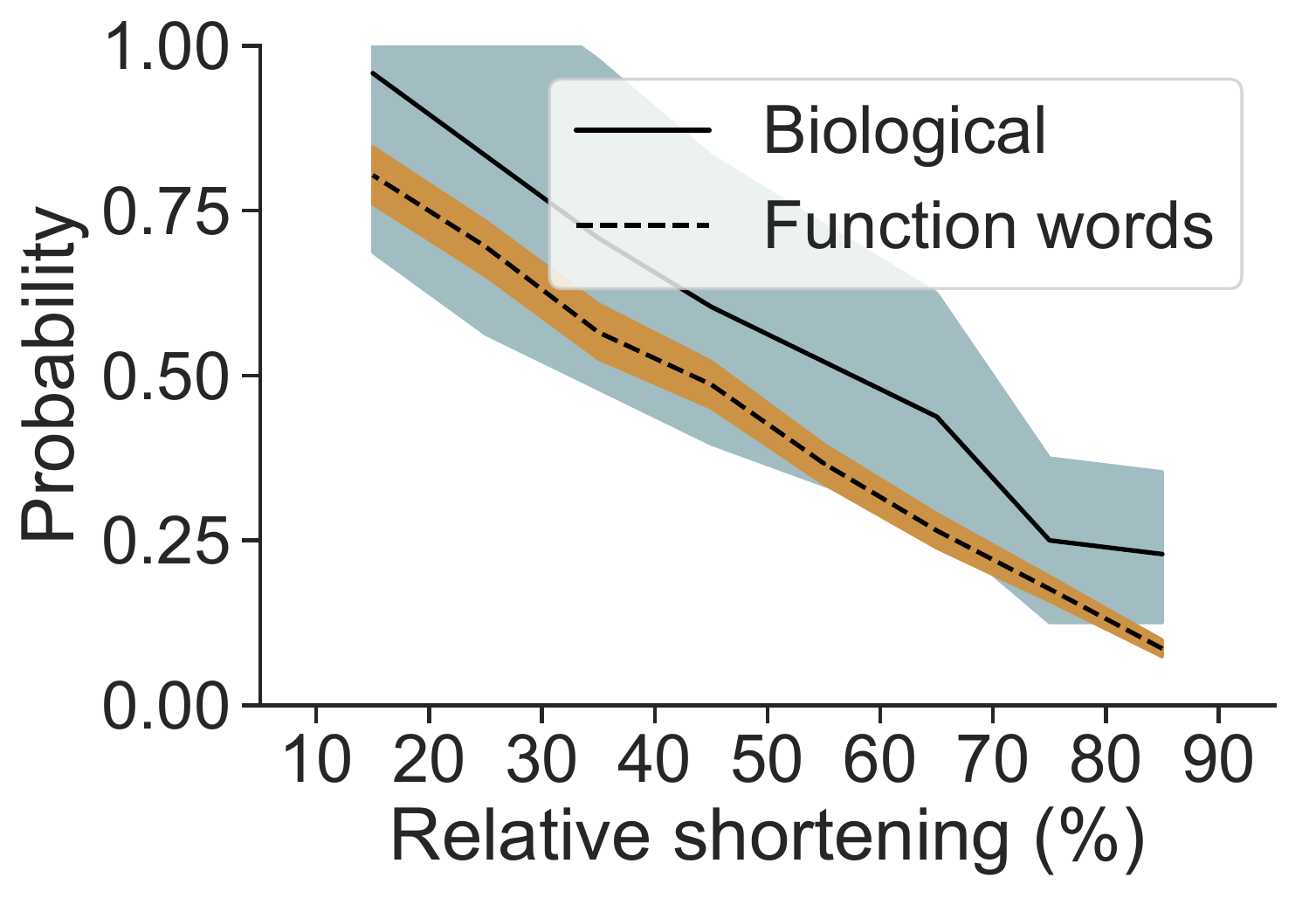}
        \subcaption{Biological*}
        \label{fig:14}
    \end{minipage}
    \hfill
    \begin{minipage}{.3\textwidth}
        \centering
        \includegraphics[width=\textwidth]{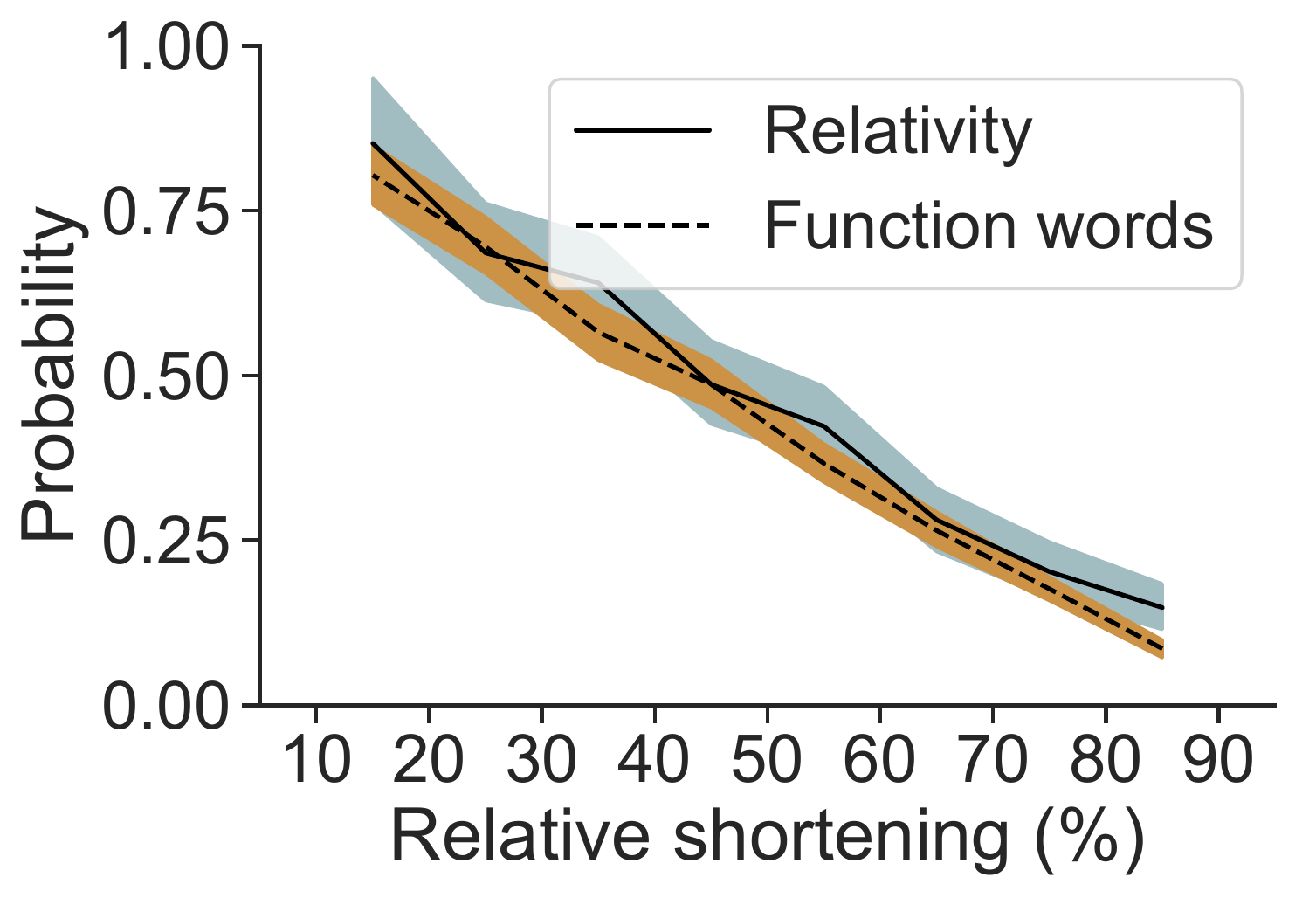}
        \subcaption{Relativity}
        \label{fig:16}
    \end{minipage}
    \caption{
    Probability of being preserved for tokens carrying psychological processes, with 95\% bootstrapped confidence intervals. The probability is compared to the baseline trend for function words. Significant differences ($p<0.0085$) are marked with an asterisk (*). Words describing affects and perceptual words are maintained more than function words. On the other hand, words describing cognitive processes and relativity are preserved less.}
    \label{fig:15}
\end{figure}

\xhdr{Psychological aspects} Finally, we analyze tweets at the content level. In particular, we study the probability of being preserved for tokens that capture psychological processes as defined by LIWC \cite{pennebaker2001linguistic}. In \Figref{fig:15}, we compare the probabilities for such tokens across different levels of length reduction. We compare the trend for each feature against the same baseline --- function words --- as they do not carry meaning of their own. We report significance levels according to Chow's test (\cf\ \Secref{sec:results}) using the adjusted significance level of $\alpha = 0.0085$, due to multiple hypothesis testing for the six psychological processes.
We note that in the process of editing a tweet to make it more concise, words describing affects (\eg, \textit{nice, love, nasty}) are preserved more than function words ($p<10^{-7}$). On the other hand, words describing cognitive processes (such as \textit{know, think, maybe}) and relativity (\textit{end, until, down, in}) are preserved less (no significant differences compared to the baseline). Perceptual words (\textit{look, feel, hear}) have a better chance of being preserved compared to the function-word baseline ($p<10^{-4}$). Finally, words that carry affect (\Figref{fig:10}) or describe subjective perceptions (\Figref{fig:13}) are preserved disproportionately more than cognitive words that indicate reasoning (\Figref{fig:12}).

\xhdr{Positive \vs\ negative affect} 
To conclude this section, we study the tokens' probability of being preserved (averaged over different levels of shortening) for subtypes of affect. In general, tokens carrying affect are likely to be kept (\Figref{fig:15}). Interestingly, the effect is much stronger for negative emotions than for positive emotions. The trend is consistent across subtypes of negative affect (anger, sadness, and anxiety. As shown in Table \ref{tab:4}), the results indicate that workers perceived negative emotions as being more important when trying to convey the meaning of the message using fewer characters.

\section{RQ3: Effectiveness of shortening strategies}
\label{sec:languageb}

For our final analysis, we observe the different editing strategies workers used to shorten tweets and seek to understand their associations with success. 

\xhdr{Extractiveness} Given the similarity of tweet shortening with the classic natural language processing task of sentence summarization, we investigate to what extent metrics used to evaluate extractive summaries, \eg, Rouge \cite{rouge}, can predict success. We observe that there is no significant correlation between unigram- and bigram-based Rouge and the probability of success at any level of shortening ($p > 0.05$), indicating that extractiveness is not the mechanism driving the observed effect.
Rather, it is the above\hyp described lexical and semantic properties that drive success.

\begin{table}[b]
  \caption{Effective and ineffective tweet\hyp shortening strategies. For each strategy, for a given token, \textit{N edits} represents the number of instances in which the insertion or deletion was made. \textit{N successful edits} (\textit{N unsuccessful edits}) is the number of times the shortened tweets associated with the edit were better (worse) than the baseline\hyp level edit. Each edit has an associated effect strength and $p$-value.}
  \label{tab:5}
\begin{tabular}{lllll}
\toprule
\multicolumn{5}{l}{\textbf{Effective strategies}} \\
\midrule
Token & N edits & N successful edits  & Effect strength & $p$-value \\
\midrule
\multicolumn{5}{l}{\textbf{Deleting}}\\
any & 36 & 24 & 3.71\% & 0.02 \\
few & 11 & 11 & 17.06\% & 0.0005 \\
more & 28 & 19 & 6.84\% &0.03 \\
a  & 285 & 161 & 2.92\%  & 0.005  \\
therefore & 23 & 17 & 6.74\% & 0.02 \\
so  & 122 & 72  & 3.03\% & 0.02\\
\multicolumn{5}{l}{\textbf{Inserting}}  \\
, & 139 & 81 & 4.18\% & 0.03\\
. & 502 & 277 & 1.23\% & 0.01\\
\midrule

\multicolumn{5}{l}{\textbf{Ineffective strategies}}\\
\midrule
Token & N edits & N unsuccessful edits  & Effect strength & $p$-value    \\
\midrule
\multicolumn{5}{l}{\textbf{Deleting}} \\
hashtags & 71 & 53 & 7.51\% & 0.00002 \\
 ? & 34 & 25 & 9.03\%& 0.005 \\
 ! & 88 & 61 & 7.58\%& 0.0002\\

 \end{tabular}
\end{table}

\xhdr{Effective \vs\ ineffective shortening strategies}
Using the same methods as those described in \Secref{sec:languagea}, we now explicitly track each token present in original and concise tweets to detect operations of deletion, insertion, and editing, with the goal of understanding the implications of these edits on the chances of success of the final, edited version over the original version of the tweet.

Since each tweet is edited a single time for a given level of reduction, we cannot make comparisons controlling for the initial tweet and the level of reduction simultaneously. As the probability of success varies more across limits than it does across tweets, we compare the success of the edited tweet with the success of the median edited tweet at the respective level of reduction. We refer to this median tweet as the baseline\hyp level edit.

For each word and each edit operation (insertion or deletion), we calculate the frequency with which the edited version was better (or worse) than the level\hyp baseline edit. We test whether the frequency is significantly different from 50\%, using a one-sided binomial test to reject the null hypothesis that the probability of success in a Bernoulli experiment is 50\%.
We also report the \textit{effect strength} for each edit operation, \ie, the average difference in the probability of success between the level\hyp baseline edit and the observed instance containing the edit.

\Tabref{tab:5} reveals that successful editing strategies include omitting certain quantifiers (\textit{any} and \textit{few}), articles (\textit{a}), and linkers (\textit{so} and \textit{therefore}), as they do not carry essential information. Deleting them is an efficient way to save space. Inserting commas and full-stops is also effective. This insertion is associated with splitting long sentences to increase readability.

Ineffective editing strategies are deleting hashtags (such as \#cyberpsa, \#gonetopot, and \#maga). Hashtags increase the visibility of tweets and make it possible for the content to potentially reach more people. It is also ineffective to delete question and exclamation marks, expressive parts of the text that can elicit discussion and reactions.

\section{Discussion and conclusions}
\label{sec:discussion}
In this work, our goals are threefold: to measure the effects of length constraints on tweet quality, to determine the linguistic traits of brevity, and to find when brevity is beneficial. To address these goals, we designed a large experiment and deployed it on the Amazon Mechanical Turk crowdsourcing platform. Our two-stage experimental pipeline first generated valid, shortened versions of original tweets, then measured the quality of the shortened versions compared with the originals. In our experiment, we observed that tweets of an original length of 250 characters can be successfully reduced by up to 30--40\% of their original length with no reduction in quality. The optimal range of shortening is consistently reducing between 10\% and 20\% of the original length.

Our results reveal a causal mechanism through which enforcing brevity during the content production process improves the content's perceived success. The initial question could also conceivably be framed the other way around: does asking workers to improve a tweet lead to the tweet being shortened? Instructing distinct crowd workers to improve a small subset of the original input tweets with the goal of making the tweet more retweeted, we observed that the workers chose to shorten the tweets in most cases, by a median of 16\%.

We also discover that concise tweets have distinctive linguistic features. In the shortening process, verbs and negations survive the most, consistent with them being the parts of speech that convey essential information, while articles, adverbs, and conjunctions have the highest probability of being omitted. Also, shortened tweets preserve the original affect and subjective perceptions surprisingly well. This is particularly the case for all subtypes of negative affect (anger, sadness, and anxiety), confirming the general principle across a broad range of psychological phenomena that negative emotions have more impact than positive emotions, and that negative information is processed more thoroughly than positive information~\cite{baumeister2001bad}. These results, especially our finding that length constraints disproportionately preserve negative emotions, have immediate implications for social media content.

We associate various editing strategies with our measure of success to try to tell successful from unsuccessful strategies. We find that, for example, successful editing strategies include omitting certain quantifiers, articles and linkers, parts of speech that do not carry essential information, and inserting full-stops and commas to break up long sentences. We note that these insights are aligned with qualitative results pointing to simplicity, clarity and directness as desirable features of tweets (\Figref{fig:reponses}). Similarly, ineffective strategies are deleting hashtags, question and exclamation marks, a result referring back to the potential to initiate discussions, a frequently occurring justification when preferring control tweets. We emphasize that this analysis is suggestive and deserves further study, given that it was not performed as a randomized experiment (unlike our main analysis). 

Our results show that imposing length constraints has a heterogeneous effect depending on the original message, as presented in \Figref{fig:2}. Although shortening by 10-20\% is optimal on average, many tweets see the largest probability of success in more drastic shortening, and some in no shortening at all. We hypothesize that at least part of this effect is driven by the fact that different messages have different potential to be edited---\ie, they differ in how ``compressed'' they are to begin with, and thus how ``compressible'' they still are.

Finally, although the isolated effect of brevity on message success is on average positive, brevity can have its disadvantages. As revealed through qualitative analysis in \Figref{fig:reponses}, brevity is not desirable when it comes at the cost of delivering a less informative or complete message. Similarly, a prevalent explanation for aversion towards brevity is a lack of emotionality. Although the shortening process strives to maintain affect (\Figref{fig:15}), this becomes increasingly tricky as the allowed length decreases. At times, even though preserving the completeness of the information is possible, being wordy and less direct is the social norm. Lacking adherence to the norm by being abrupt might be perceived as impolite, and therefore socially undesirable \cite{locher2008relational, danescu2013computational}.

Our results have implications for the design of online social media platforms, many of which enforce certain length restrictions for posts. Length constraints might push users towards editing their content in a useful way and influence users to create concise content that other users are more likely to engage with. However, our work finds that the very design of social media platforms promotes negativity in unintended ways.

\xhdr{Limitations and future work} An interesting direction for future work is to further illuminate, and perhaps predict, how compressible a message is. With a larger corpus of tweets, it might be possible to infer an optimal level of shortening for a given message.

We note that, despite our best effort, our experimental setup may not fully guarantee that the semantic content of tweets is entirely preserved in the process of shortening. The comprehension questions might not perfectly capture all the information, or the workers performing the shortening validation task (Task~4) might guess the correct answers. In the example presented in \Tabref{tab:2}, even though the tweet shortened to 30--40\% of the original length is more successful compared to the original tweet, it does not contain the information that addiction is difficult to handle. We note that despite this limitation, our experiment is still far more controlled compared to the observational approaches previously used to address our research questions.

Our experimental setup is designed to replicate phenomena occurring on Twitter. As discussed in \Secref{sec:rel}, the fact that crowd workers are accurate in predicting success on Twitter brings validity to our experimental design \cite{r9}, although we note that crowdsourced ratings are only a proxy for actual perception of success on social media.
Future work should study the extent to which crowdsourced ratings evaluate the expected popularity of the content as opposed to reflecting the participants' own biases about what other people would be likely to engage with \cite{Thebault-Spieker:2017:SER:3171581.3134736}.

Additionally, we cannot guarantee our findings generalize beyond Twitter. Although it is reasonable to assume that similar cognitive mechanisms govern the observed effects of brevity on message success in general, future work should explore to what extent brevity is beneficial on different platforms and for different content.
Future studies should evaluate the effects of brevity on a larger sample of tweets and on a sample of tweets written by highly influential users producing content of atypically high public appeal. Similarly, future studies should also measure the upper bounds to the benefits of brevity when shortening is performed by highly skilled authors.

Even though we discuss general linguistic and semantic features, such as verbs or affect, rather than English\hyp specific characteristics, we currently cannot generalize our findings to other languages. Some languages might be linguistically denser than others, and cultural aspects might affect the wording in complex ways. Future work should therefore generalize our study to multilingual settings in order to discover the effects of conciseness beyond English.

\begin{acks}
We would like to thank Sid Suri, Manoel Horta Ribeiro, Tiziano Piccardi, and our reviewers for their helpful suggestions. This research has been supported in part by Microsoft, Google, CROSS, and NSERC. 
\end{acks}

\bibliographystyle{ACM-Reference-Format}
\bibliography{bibliography}

\end{document}